\documentclass[11pt,a4paper,english,twoside]{article}

\usepackage{a4wide}
\usepackage{amssymb, amsmath}
\usepackage{graphicx}
\usepackage[all]{xy}
\usepackage{hyperref}
\hypersetup{linktocpage}
\usepackage{enumerate}
\usepackage{xcolor}
\usepackage{dsfont}
\usepackage{empheq}
\usepackage{cite}
\usepackage{float}

\usepackage[utf8]{inputenc}
\usepackage{soul}
\usepackage{subcaption}
\usepackage{hhline}

\newcommand{\beq}{\begin{equation}}
\newcommand{\eeq}{\end{equation}}
\def\bea#1\eea{\begin{align}#1\end{align}}
\def\beal#1\eeal{\begin{subequations}\begin{align}#1\end{align}\end{subequations}}
\newcommand{\nn}{\nonumber}
\newcommand{\w}{\wedge}
\renewcommand{\i}{\ensuremath{\textnormal{i}}}

\newcommand{\eq}[1]{\begin{equation}#1\end{equation}}
\newcommand{\spl}[1]{\begin{split}#1\end{split}}

\def\del {\partial}
\def\d {{\rm d}}

\def\mcal{\mathcal{M}}

\def\ocal{\mathcal{O}}

\begin{document}
\numberwithin{equation}{section}

\begin{titlepage}

\begin{center}

\phantom{DRAFT}

\vspace{2.4cm}

{\LARGE \bf{Gravitational waves in warped compactifications}}\\

\vspace{2.2 cm} {\Large David Andriot$^{1,2}$ and Dimitrios Tsimpis$^{3}$}\\
 \vspace{0.9 cm} {\small\slshape $^1$ Institute for Theoretical Physics, TU Wien\\
Wiedner Hauptstrasse 8-10/136, A-1040 Vienna, Austria}\\
 \vspace{0.2 cm} {\small\slshape $^2$ Theoretical Physics Department, CERN\\
1211 Geneva 23, Switzerland}\\
 \vspace{0.2 cm} {\small\slshape $^3$ Institut de Physique des Deux Infinis de Lyon\\
Universit\'{e} de Lyon, UCBL, UMR 5822, CNRS/IN2P3\\
4 rue Enrico Fermi, 69622 Villeurbanne Cedex, France}\\
\vspace{0.5cm} {\upshape\ttfamily david.andriot@tuwien.ac.at; tsimpis@ipnl.in2p3.fr}\\

\vspace{2.8cm}

{\bf Abstract}
\vspace{0.1cm}
\end{center}

\begin{quotation}
\noindent
We study gravitational waves propagating on a warped Minkowski space-time with $D-4$ compact extra dimensions. While Kaluza--Klein scales are typically too high for any current detection, we analyse how the warp factor changes the Kaluza--Klein spectrum of gravitational waves. To that end we provide a complete and explicit expression for the warp factor, as well as the Green's function, on a $d$-dimensional torus. This expression differs from that of braneworld models and should find further uses in string compactifications. We then evaluate the Kaluza--Klein spectrum of gravitational waves. Our preliminary numerical results indicate not only a deviation from the standard toroidal spectrum, but also that the first masses get lowered due to the warp factor.

\end{quotation}

\end{titlepage}

\newpage

\noindent\rule[1ex]{\textwidth}{1pt}
\vspace{-0.8cm}

\tableofcontents

\vspace{0.5cm}
\noindent\rule[1ex]{\textwidth}{1pt}
\vspace{0.3cm}

\section{Introduction}

The observation of gravitational waves of astrophysical origin by the LIGO and Virgo collaboration \cite{LIGOScientific:2018mvr} has opened a new era where we have at hand an independent experimental device to scrutinize nature and test our theories. The prospects of gravitational wave observations are flourishing: much is expected from the advanced LIGO and Virgo, but also from the upcoming KAGRA, IndIGO, SKA, eLISA, as well as the awaited results of PTAs, BICEP3, etc. It is then legitimate to ask ourselves whether new fundamental physics can be probed through these observations \cite{Yunes:2016jcc, Ezquiaga:2017ekz, Barack:2018yly, Ezquiaga:2018btd, Nair:2019iur, Calcagni:2019ngc}. Here, we are interested in the question of extra dimensions. Consider gravitational waves that have been emitted in a $D$-dimensional space-time, and now propagate away from the source. A generic choice for the propagation background is then a warped four-dimensional (4d) maximally symmetric space-time, e.g.~Minkowski, parameterized by $x^{\mu}$, together with $D-4$ extra dimensions  parameterized by $y^p$, with metric
\eq{\label{metricintro}
\d s^2_{D}=e^{2A(y)}\ \tilde{g}_{\mu\nu}\d x^{\mu}\d x^{\nu}+ \hat{g}_{pq} \d y^p \d y^q
~,}
where the warp factor $H(y)$ is a power of $e^{2A}$ which depends on the setup. In this framework, three types of effects due to extra dimensions on 4d gravitational waves can be identified \cite{Andriot:2017oaz}. First, gravitational waves along the extra dimensions can appear in 4d as extra vector and scalar fields, which can couple and affect the 4d gravitational waves; in particular, a corresponding massless scalar field was shown in \cite{Andriot:2017oaz} to affect the polarization.\footnote{Since then, interesting prospects on constraints from polarization were discussed in \cite{Isi:2017fbj, Hagihara:2019ihn}, while observational constraints remained so far very weak \cite{Abbott:2017oio, Abbott:2018lct}. By providing a light mass to the scalar, further interesting setups and constraints were discussed in e.g.~\cite{Brito:2017zvb, Grin:2019mub}, while a possibility for scalar waves emission was pointed-out in \cite{Wong:2019yoc}.} Secondly, the dependence on the extra coordinates leads to Kaluza--Klein towers of fields. However, in standard models, the Kaluza--Klein 4d gravitational waves have a mass or frequency too high to be detected with current instruments: the observational upper bound of $10^{-4}$ m (about $10^{-3}$ eV) on the size of an extra dimension leads to a frequency $10^8$ times bigger than the LIGO sensitivity bound \cite{Andriot:2017oaz, Cardoso:2019vof}. Thirdly, the warp factor allows for new couplings and leads to further modifications. In this paper, we are interested in the last two effects: in particular, what is the impact of the warp factor on the Kaluza--Klein spectrum? We want to evaluate deviations from the standard spectrum due to the warp factor,\footnote{\label{foot:intro}The spectrum will be compared to both the Kaluza--Klein spectrum for a constant warp factor, and the one corresponding to an average (warped) internal length; both will turn out to be the same here. The deviation and lowering of the spectrum will then be due to the non-constant part of the warp factor.} and see whether its contribution could lower  the Kaluza--Klein masses enough, so that 4d Kaluza--Klein gravitational waves become observable. Most likely, the contribution will not be enough to reach LIGO frequency range, but it could be interesting for high energy primordial gravitational waves to be detected by eLISA, as discussed in section \ref{sec:ccl}.

Models with warped extra dimensions are numerous, ranging from braneworld models {\`a} la Randall-Sundrum \cite{Randall:1999ee, Randall:1999vf} to string compactifications \cite{Becker:1996gj, Dasgupta:1999ss, Giddings:2001yu}. Generally, the warp factor accounts for the backreaction of an extended object like a brane, and typically its singular locus coincides with the position of the object in its transverse space. Expressions for the warp factor in braneworld models or stringy frameworks are however very different. For the former, the warp factor is usually explicitly given, typically an exponential of an extra dimension coordinate; its impact on gravitational waves has been discussed in e.g.~\cite{Davoudiasl:1999jd, Seahra:2004fg, Clarkson:2006pq, Dillon:2016fgw, Chakraborty:2017qve, Visinelli:2017bny}. For stringy models, the warp factor usually comes from $p$-brane solutions: these solutions of supergravity-like theories, describing extended objects, connect to $D_p$-branes in string theory. The warp factor is defined as the solution to a Poisson equation, that relates it to Green's functions. For non-compact extra dimensions, Green's functions are usually well-known and so is the warp factor. It is however not the case for compact extra dimensions with branes, despite the fact that all these ingredients are appealing for realistic phenomenology. As a consequence, in explicit string compactifications towards 4d Minkowski \cite{Giddings:2001yu, Andriot:2016ufg}, the warp factor is not given: its defining equation is known, \eqref{BI}, but not solved explicitly, even though there is no doubt on the existence of a solution (see also \cite{Andriot:2015sia, Macpherson:2016xwk, Apruzzi:2018cvq}). The warp factor on compact spaces, and the throats it generates, however play a crucial role in many aspects of (string) phenomenology, as for instance in the recent developments regarding de Sitter solutions \cite{Bena:2018fqc, Cordova:2018dbb, Carta:2019rhx, Blumenhagen:2019qcg, Cribiori:2019clo, Bena:2019sxm}; we come back to these applications in section \ref{sec:ccl}. An intermediate task in this paper is therefore to determine explicitly the (stringy or $p$-brane) warp factor for compact extra dimensions, and we will do so for a transverse $d$-dimensional torus $\mathbb{T}^d$.

We first provide in section \ref{sec:KKpbrane} all relevant equations to determine the Kaluza--Klein spectrum of gravitational waves propagating on a 4d warped Minkowski background with extra dimensions. We follow \cite{Bachas:2011xa} and \cite{Andriot:2017oaz} to obtain the eigenvalue equation that gives the spectrum: it generalizes the standard Laplacian equation with a dependence on the warp factor. We further recall $p$-brane solutions in $D$-dimensional space-times for non-compact or compact extra dimensions, and provide the defining equation for the warp factor $H$. A solution to the latter is then given in terms of generalized Green's functions, the source charges (e.g.~generic dipoles, or $D$-branes and orientifolds in a stringy framework), and a constant $H_0$. In section \ref{sec:Greengal}, we provide an explicit expression for generalized Green's functions on a torus $\mathbb{T}^d$, inspired by Courant-Hilbert \cite{Courant:1989aa} and recent proposals \cite{Shandera:2003gx, Kim:2018vgz}. From this expression, we reproduce analytically the expected flat space behaviour close to the source in appendix \ref{ap:behaviour}. For $d=2$, we observe in appendix \ref{ap:2d} a perhaps surprising matching with a different expression known from string amplitudes. To complete the warp factor expression, we discuss the constant $H_0$, which turns out to play a crucial role. We propose a prescription to fix it, and compute it in some cases. We finally turn to the Kaluza--Klein gravitational wave spectrum in section \ref{sec:spectrumgal}. We characterize analytically what choices of parameters (the constant $H_0$, the size of $\mathbb{T}^d$, etc.) would lead to a deviation from the standard Kaluza--Klein spectrum without warp factor (i.e.~a constant one), while staying in a physically relevant regime (e.g.~typical lengths bigger than the fundamental length $l_s$). We then evaluate numerically the Kaluza--Klein spectrum with the above warp factor generated by $D$-branes and orientifolds, or analogous sources in $D$ dimensions. We do so for $d=1,2,3$ transverse dimensions, with different values of the parameters. For $d=1$, this analysis is completed in appendix \ref{ap:num} by an alternative method, used already in \cite{Richard:2014qsa}. The spectrum evaluation quantifies the deviations from the standard Kaluza--Klein spectrum. We summarize and discuss the results in section \ref{sec:ccl}, in particular the observability of these deviations.

\section{Kaluza-Klein gravitational waves and $p$-brane backgrounds}\label{sec:KKpbrane}

In this section, we first introduce four-dimensional Kaluza--Klein gravitational waves on a warped Minkowski background with extra dimensions. We identify the eigenfunction equation relevant to determine their spectrum. We then present the $p$-brane solutions that will serve as the background. We consider non-compact or compact extra dimensions, and discuss the related definitions of the warp factor. We first give them in a $D=10$ string context, and then more generally in arbitrary $D$ dimensions. We finally rewrite the eigenfunction (spectrum) equation on such backgrounds with toroidal extra dimensions.

\subsection{Spectrum for a warped background}\label{sec:1mg}

We consider the following metric of a $D$-dimensional space-time
\eq{\label{metricpert}\d s^2_{D}=
e^{2A}\left(\eta_{\mu\nu}+h_{\mu\nu}\right)\d x^{\mu}\d x^{\nu}+ \hat{g}_{pq} \d y^p \d y^q
~,}
where $x^{\mu=0,\dots, 3}$ are coordinates of $\mathbb{R}^{1,3}$, $y^{p}$ those of the $(D-4)$-dimensional ``internal'' space $\mcal$ with metric $\hat{g}_{pq}$, and the warp factor (a power of $e^A$) depending only on the $y^{p}$. Over the background metric $e^{2A} \eta_{\mu\nu}$ and $\hat{g}_{pq}$, we consider four-dimensional (4d) metric perturbations $h_{\mu\nu}(x,y)$. Equations describing their dynamics at linear order were obtained in \cite{Bachas:2011xa}, generalizing \cite{Csaki:2000fc} (see also \cite{Andriot:2017oaz}). First, these perturbations can be decomposed into a tower of Kaluza--Klein modes
\eq{\label{f}
h_{\mu\nu}(x,y)=\sum_N h^{N}_{\mu\nu}(x^{\mu}) \, \psi_N(y^p)
~.}
The $\psi_N$'s are orthonormal (weighted) eigenfunctions of a modified Laplacian of $\mcal$ with eigenvalues $M_N^2$:
\eq{\label{int}
-\frac{1}{\sqrt{\hat{g}}}\partial_p \left(\sqrt{\hat{g}}\, \hat{g}^{pq}e^{4A} \, \partial_q \psi_N \right)=M_N^2\, e^{2A}\, \psi_N
~,}
with $\hat{g}$ the determinant of $\hat{g}_{pq}$. The Kaluza--Klein modes $h^{N}_{\mu\nu}(x)$ in the expansion \eqref{f} are taken transverse and traceless (TT) in 4d, i.e.~$\eta^{\kappa\lambda}\partial_{\kappa}h^{N}_{\lambda\mu}=0, \, \eta^{\mu\nu}h^{N}_{\mu\nu}=0$. Then they obey the Pauli-Fierz equation for a massive spin-2 field of mass $M_N$ in $\mathbb{R}^{1,3}$
\eq{\label{pf}\left(\eta^{\kappa\lambda}\partial_{\kappa}\partial_{\lambda} -M_N^2 \right)h^{N}_{\mu\nu} = 0
~.}
Indeed, provided the unperturbed or background metric is a solution of the $D$-dimensional theory (in Einstein frame), \cite{Bachas:2011xa} shows that the $D$-dimensional linearized Einstein equations are satisfied as just described. Interestingly, this is argued to hold {\it independently} of the form of the energy-momentum tensor for the ``matter'' fields (i.e.~all  fields other than the metric). This is because one considers a maximally symmetric 4d space-time, thus constraining possible contributions to the matter energy momentum tensor. The above equations are then relevant to describe the propagation of 4d (Kaluza--Klein) gravitational waves on a warped Minkowski background with $D-4$ extra dimensions, for any matter content of the theory; this is the question of interest here. Physically, this is an appropriate description away from the source where gravitational waves have been emitted.

As a side a remark, note that a more general analysis was made in \cite{Andriot:2017oaz} by allowing further metric fluctuations $h_{pq}$ and $h_{\mu q}$. As shown  in appendix A.2
therein, setting $h_{pq}=h_{\mu q}=0$ and choosing the 4d TT gauge for $h_{\mu\nu}$ makes the $D$-dimensional equations reduce to those above. In other words, the fluctuations considered above are an ansatz that provides a consistent truncation of the fully fluctuated $D$-dimensional theory, meaning that solutions to the above would also be (linear) solutions to the $D$-dimensional theory with all metric fluctuations. This point could be of interest for extensions of the present work.

To have normalisable spin-2 excitations, the corresponding eigenmodes $\psi_N$ of the modified Laplacian \eqref{int} should verify
\eq{\label{norm}\int_{\mcal}\d^{D-4}y \sqrt{\hat{g}}\, e^{2A}\, |\psi_N|^2<\infty~.}
Moreover \cite{Bachas:2011xa} shows that $M_N^2\geq 0$, with the lower bound saturated, $M_0=0$, if and only if the corresponding eigenmode is constant: $\psi_0=\mathrm{const}$. This mode corresponds physically to the standard (massless) 4d gravitational wave. Requiring it to be present gives in turn a constraint on the warp factor, by imposing the finiteness of the integral \eqref{norm} with a constant $\psi_0$: we come back to this constraint in section \ref{sec:normH}. To go further, we now need to specify the background metric, and in particular determine the warp factor. Warp factors typically account for the backreaction of extended objects like branes, so we turn  to $p$-brane solutions in the next section.

\subsection{$p$-branes in $D=10$}\label{sec:brane10}

The $p$-brane solutions exist in any dimension $D$ (see e.g.~\cite{Duff:1993ye} and section 6.1 of \cite{Youm:1997hw} for a review). We first present them here in $D=10$, where their relation to $D_p$-branes of string theory can most easily be established, providing precise values for their tension and charge and fixing our conventions. We also emphasize differences between non-compact or compact extra dimensions. In the next section, we briefly generalize to any $D$, and use such backgrounds for our original gravitational waves problem.

The $p$-brane solutions in $D=10$ are most easily expressed in string frame: they are then solutions to the action $S=S_{{\rm bulk}} + S_{{\rm sources}}$ where
\beq
S_{{\rm bulk}} = \frac{1}{2\kappa_{10}^2} \int \d^{10} x \, \sqrt{|g_{10}|}\ \left( e^{-2\phi} ({\cal R} +4 |\d \phi|^2 ) - \frac{1}{2} |F_{p+2}|^2 \right)\ ,
\eeq
with $g_{10}$ the determinant of the 10d metric $g_{MN}$, $\phi$ the dilaton, and the abelian $(p+2)$-form field strength $F_{p+2} = \d C_{p+1}$. The square of a $q$-form $A_q$ is $|A_q|^2 = A_{q\, M_1 \dots M_q}\, A_{q}^{\ \, M_1 \dots M_q} / q!$, lifting indices with $g^{MN}$. In addition, we have
\beq
S_{{\rm sources}} = - T_p \int_{\Sigma_{p+1}} \d^{p+1}\xi \ e^{-\phi} \sqrt{|\imath^*[g_D]|} + \mu_p \int_{\Sigma_{p+1}} \imath^*[C_{p+1}] \ ,
\eeq
where $\Sigma_{p+1}$ is the source world-volume with coordinates $\xi^{i=0 \dots p}$, and $\imath^*[ \cdot]$ the pull-back to it. These actions can be promoted, at least for trivial embedding, to 10d ones including appropriate $\delta$-functions over the transverse space (see e.g.~conventions in appendix A of \cite{Andriot:2016xvq}). The gravitational constant and the tension are given in $D=10$ by
\beq
2\kappa_{10}^2=(2\pi)^7 (\alpha^\prime)^4\ , \quad T_p^2=\frac{\pi}{\kappa_{10}^2} (4\pi^2 \alpha^\prime)^{3-p} \ ,
\eeq
where we take $\alpha^\prime=l_s^2$ with string length $l_s$. For BPS sources as here, one has for the charge $\mu_p=T_p$. The $p$-brane solutions in $D=10$ in string frame are then given by (see e.g.~\cite{Johnson:2000ch})
\eq{\spl{\label{solgenstring}
\d s^2&= H^{-\frac{1}{2}} \eta_{ij}\d x^i\d x^j+ H^{\frac{1}{2}} \delta_{mn}\d y^m\d y^n\\
e^\phi&=e^{\phi_0}\, H^{-\frac{p-3}{4}}~;~~~C_{{p+1}}=(H^{-1} -1)\, e^{-\phi_0}\, \text{vol}_{p+1}
~,}}
with $x^{i=0 \dots p}$ for the parallel space with $\text{vol}_{p+1} = \d x^0 \w \dots \w \d x^p$, and $y^{m=p+1 \dots 9}$ for the transverse space. The constant $e^{\phi_0}=g_s$ determines the string coupling. The function $H(\vec{y})$ is what we call the warp factor; it obeys
\eq{\label{laplpbrane}
\delta^{mn}\partial_m\partial_n H(\vec{y})= \tilde{Q}\, \delta(\vec{y}-\vec{y}_0)\ , \quad \tilde{Q} \propto g_s \kappa^2_{10} T_p
~,}
where the brane is located thanks to the Dirac $\delta$-function at $\vec{y}_0$ in the (unwarped) transverse space $\mathbb{R}^{9-p}$, and the numerical factor in the charge $\tilde{Q}$ will be specified below. In that case where the extra dimensions, especially the transverse space, are non-compact, $H$ is a well-known Green's function in flat space.

These solutions find generalizations in string flux compactifications, where the extra dimensions are compact. In $D=10$ type II supergravities, the solution of \cite{Giddings:2001yu} for $p=3$ got generalized to a large class of Minkowski solutions in \cite{Andriot:2016ufg}, where the metric in 10d string frame is given by
\beq
\d s^2_{10}= H^{-\frac{1}{2}} ( \eta_{\mu \nu}\d x^{\mu}\d x^{\nu} + \d \tilde{s}^2_{6||} ) + H^{\frac{1}{2}} \d \tilde{s}^2_{6\bot} \ . \label{10dmetricstringframe}
\eeq
This is the same as the $p$-brane \eqref{solgenstring} except that the 6 extra dimensions are now compact, and so far not necessarily flat. In these solutions one has again $e^{\phi} = g_s H^{-\frac{p-3}{4}}$, and a similar expression for the sourced flux $F_{p+2}$, that gives rise to the following defining equation for the warp factor
\beq
\Delta H + C = \frac{1}{\sqrt{g}} \sum_i Q_i\, \delta(\vec{y} - \vec{y}_i) \ .\label{BI}
\eeq
This involves analogous ingredients to the non-compact counterpart \eqref{laplpbrane}: $\Delta$ is the Laplacian on the (unwarped, compact) transverse space to the sources, of dimension $d=9-p$, with metric determinant denoted $g$; $Q_i$ are charges due to branes $D_p$ or orientifolds $O_p$ localised by the $\delta$-functions at $\vec{y}_i$; $C$ is a positive quantity, typically a constant due to extra fluxes or internal curvature. Further needed conventions are as follows: the volume form on the (unwarped) transverse space is denoted ${\rm vol}_{\bot}$, and we use the shorthand notation $V=\int {\rm vol}_{\bot}$ for the volume. We take here as a convention $\int \d^{9-p} y\, \delta(\vec{y} - \vec{y}_i) = \int \frac{{\rm vol}_{\bot}}{\sqrt{g}}\, \delta(\vec{y} - \vec{y}_i) = 1$. The function $H$ is usually considered well-behaved enough such that the integral of \eqref{BI} on the compact space gives $C = \tfrac{1}{V} \sum_i Q_i $. Finally, we have
\beq
\label{52}
Q_{D_p}= - 2 \kappa^2_{10} T_p g_s = -(2\pi l_s)^{7-p} g_s \ , \quad Q_{O_p}=- 2^{p-5} Q_{D_p} \ .
\eeq

While in the non-compact case, solutions to \eqref{laplpbrane} are well-known Green's functions in flat space, $H$ in the compact case is most of the time undetermined, e.g.~in string compactifications. A first idea could be to start from the non-compact situation and identify periodically the transverse $y$-coordinates to get a torus. It is however easily seen that in a compact case, \eqref{laplpbrane} cannot be solved for a non-vanishing $\tilde{Q}$, by integrating both sides of that equation. This is because on a compact space without boundary, the flux of a point charge has nowhere to go. So one needs at least an opposite charge, and adding it in \eqref{laplpbrane}, leads essentially to \eqref{BI}. In a string theory context, it is well-known \cite{Maldacena:2000mw} that due to compactness, $O_p$ need to be  included to compensate the charges of the $D_p$, hence the sum over sources in \eqref{BI}. However, a Green's function is defined with only one $\delta$-function, so one introduces a so-called generalized Green's function $G$ that solves
\begin{empheq}[innerbox=\fbox]{align}
\Delta G (\vec{y}) & =\frac{\delta (\vec{y})}{\sqrt{g}}  - \frac{1}{V}  \label{BIG}
\end{empheq}
On a compact space, one can integrate to zero both sides thanks to the extra constant on the right-hand side, contrary to the standard Poisson equation. The need for this extra background charge can also be understood from the fact that the Laplacian on a compact space without boundary can only be inverted on the space of functions orthogonal to the constant mode (which is the eigenfunction of the Laplacian corresponding to zero eigenvalue). In terms of such generalized Green's function, a solution to \eqref{BI} for the warp factor is then given by
\begin{empheq}[innerbox=\fbox]{align}
H & = \sum_i Q_i\, G(\vec{y} - \vec{y}_i) + H_0 \label{H}
\end{empheq}
with some constant $H_0$. We come back to the problem of determining the generalized Green's function $G$ and the constant $H_0$ for the compact case in section \ref{sec:Greengal}.

\subsection{General $D$ and gravitational wave background}\label{sec:braneD}

The $p$-brane solutions just presented can serve as a Minkowski background with warp factor, to study the propagation of gravitational waves as described in section \ref{sec:1mg}.\footnote{While it is obvious that the brane metric is appropriate for a warped background of interest, one may wonder about the additional ingredients. Concerning the latter however, we know that such a background can be used thanks to the argument of \cite{Bachas:2011xa} regarding the generic matter contribution to the energy-momentum tensor, recalled in section \ref{sec:1mg}. As a further check, an explicit computation of the matter contribution on a similar background was made in appendix B of \cite{Andriot:2017oaz}.} To that end, an extra step is to express the solution in Einstein frame, since the relevant equations in section \ref{sec:1mg} were derived there. We also generalize the previous solutions to $D$-dimensional ones.

In $D$ dimensions, the string and Einstein frames are related thanks to the change of metric $g_{MN}= e^{\frac{4(\phi - \phi_0)}{D-2}} g_{E\, MN}$, giving the relation
\beq
\frac{1}{2\kappa_{D}^2} \int \d^{D} x \, \sqrt{|g_D|}\ e^{-2\phi} \left(  {\cal R} +4 |\d \phi|^2 \right) = \frac{1}{2\kappa_{D}^2 g_s^2} \int \d^{D} x \, \sqrt{|g_E|}\ \left( {\cal R}_E -\frac{4}{D-2} |\d \phi|_E^2 \right)  \ .
\eeq
The $p$-brane metric in $D=10$ \eqref{solgenstring} becomes
\beq
\d s_E^2 = H^{\frac{p-7}{8}} \eta_{ij}\d x^i\d x^j+ H^{\frac{p+1}{8}} \delta_{mn}\d y^m\d y^n \ , \label{metricEinstein10}
\eeq
which gets generalized in $D$ dimensions \cite{Duff:1993ye} to
\beq
\d s_E^2= H^{-\frac{D-p-3}{D-2}}\eta_{ij}\d x^i \d x^j +H^{\frac{p+1}{D-2}}\delta_{mn}\d y^m\d y^n \ . \label{metricEinstein}
\eeq
We turn to the case of compact extra dimensions, and focus in the following on an unwarped transverse space being a $d$-dimensional torus, $d=D-p-1$. In the case $D=10$, the string frame metric \eqref{10dmetricstringframe} becomes in Einstein frame
\beq
\d s_E^2=  H^{\frac{p-7}{8}} ( \eta_{\mu \nu}\d x^{\mu}\d x^{\nu} + \d \tilde{s}^2_{6||} ) + H^{\frac{p+1}{8}} \delta_{mn}\d y^m\d y^n \ . \label{metricEinsteincompact10}
\eeq
If we also restrict the unwarped parallel extra dimensions to be toroidal, i.e.~$\d \tilde{s}^2_{6||}=\delta_{ij}\d x^i \d x^j$, we get (locally) the same metric as the non-compact $p$-brane one: the Einstein frame metric \eqref{metricEinstein10}, generalized to \eqref{metricEinstein} in $D$ dimensions, for that toroidal case.

We now match the above to the metric \eqref{metricpert} in the unperturbed case and get
\beq
e^{2A} = H^{-\frac{D-p-3}{D-2}} \ ,\label{warpmatch}
\eeq
and for the metric on the $(D-4)$-dimensional space $\mcal$
\beq
\hat{g}_{pq} \d y^p \d y^q :=  H^{-\frac{D-p-3}{D-2}} \delta_{ij}\d x^i \d x^j +H^{\frac{p+1}{D-2}}\delta_{mn}\d y^m\d y^n \ ,\label{MmetricEinsteinD}
\eeq
in the compact toroidal case. We then compute in this $(x^i,y^m)$ coordinate basis $\sqrt{\hat{g}}= H^{\frac{2D-5-p}{D-2}}$. The eigenfunction equation \eqref{int} remarkably simplifies to
\beq
- \delta^{ij} \partial_i \left(H \partial_j \psi_N \right) - \delta^{mn} \partial_m \partial_n \psi_N = H M_N^2\, \psi_N \ . \label{eigeneq0}
\eeq
The first term deals with the dependence on $x^i$, i.e.~parallel directions, and the second one on $y^m$, i.e.~transverse dependence. As mentioned in section \ref{sec:brane10}, $H$ only depends on the latter. The first term therefore only contributes to the eigenvalues $M_N^2$ by shifts due to $\delta^{ij} \partial_i \partial_j \psi_N$. With toroidal parallel directions, we can decompose the $x$-dependence of $\psi_N$ as Fourier series, and $M_N^2$ are then shifted by the standard discretized constants. For simplicity, we take from now on the zero-mode of that decomposition, in other words we restrict to $\psi_N(y^m)$. Upon this minor restriction, the eigenfunction equation boils down to
\begin{empheq}[innerbox=\fbox]{align}
 - \delta^{mn} \partial_m \partial_n \psi_N & = H M_N^2\, \psi_N \label{eigeneq}
\end{empheq}
where $H$ and $\psi_N$ depend only on $y^m$, coordinates of the $d$-dimensional unwarped transverse torus, with $d=D-p-1$.

Solving the previous equation will give us the desired Kaluza--Klein spectrum of gravitational waves. To that end, we need the warp factor $H$ \eqref{H}, given in terms of generalized Green's functions solving \eqref{BIG} on a $d$-dimensional torus, a constant $H_0$ and source charges. We will determine the first two in section \ref{sec:Greengal}, and we now define the last ingredient, namely the charges $Q_{D_p}$ in a general dimension $D$. Building on \eqref{52} and \cite{Duff:1993ye}, we express the $D$-dimensional physical constants naturally (using dimensional analysis and the actions) in terms of a fundamental length $l_s$ and a numerical constant $g_s$ related to $e^{\phi_0}$, as follows
\beq
2\kappa_{D}^2=\frac{(2\pi l_s)^{D-2}}{2 \pi} \ , \quad T_p^2=\frac{\pi}{\kappa_{D}^2} (2\pi l_s)^{D-4-2p} = (2 \pi)^2 (2\pi l_s)^{-2(p+1)} \ ,
\eeq
and define
\beq
Q_{D_p}= - 2 \kappa^2_{D} T_p g_s = -(2\pi l_s)^{D-p-3} g_s \ , \quad Q_{O_p}=- 2^{p-D+5} Q_{D_p} \ , \label{QD}
\eeq
where $Q_{O_p}$ is defined this way for further convenience. The interpretation of $l_s$ and $g_s$ as a string length and string coupling is most relevant in $D=10$; similarly, the $D_p$ and $O_p$ do not necessarily have the same interpretation outside of a string, $D=10$, context. For more phenomenological models, the latter can simply be viewed as two sources needed to create dipoles.

\section{Generalized Green's functions and the warp factor}\label{sec:Greengal}

To determine the warp factor $H$ \eqref{H} with compact toroidal extra dimensions, we are now interested in solving \eqref{BIG}, i.e.~identify the generalized Green's functions, on a torus $\mathbb{T}^d$. We take for simplicity the same radii $L$, i.e.~each toroidal coordinate obeys the identification $y^m \sim y^m + 2\pi L$. We introduce coordinates $\sigma^m = y^m / (2\pi L) \in [-\tfrac{1}{2},\tfrac{1}{2}]$, the torus metric is then
\beq
\d s_{\mathbb{T}^d}^2 = \delta_{mn} \d y^m \d y^n = \delta_{mn} 4 \pi^2 L^2 \d \sigma^m \d \sigma^n \ .\label{torusmetric}
\eeq
As a warmup, we consider in section \ref{sec:threebrane} a $d=3$ toy model where the warp factor is sourced by dipoles. We present an inspiring solution obtained by Courant and Hilbert. We then turn to a general solution for generalized Green's functions in section \ref{sec:Green}. We finally discuss the warp factor constant in section \ref{sec:normH}.

\subsection{Warmup: warp factor from dipoles with $d=3$}\label{sec:threebrane}

We consider branes that fill out the 4d space-time and are transverse to $d=3$ dimensions, e.g.~$p=3$ in $D=7$ or $p=6$ in $D=10$. The compact transverse space is $\mathbb{T}^3$ and we include both positive- and negative-tension objects so as to ensure that the total charge vanishes. To determine the warp factor $H$ solving \eqref{BI}, here with $C=0$ and $Q_i=\pm 1$, we follow the presentation by Courant and Hilbert \cite{Courant:1989aa} (see p.378); our $H$ is simply called there the (total) Green's function. One considers the rectangular parallelepiped bounded by the planes $y^1=\pm a/2$, $y^2=\pm b/2$ and $y^3=\pm c/2$. The aim is to compute the Green's function due to a unit charge at the point $({y}^1_0,{y}^2_0,{y}^3_0)$, as well as all image charges obtained by reflecting  the charge (with a sign flip after each reflection) across the planes of the lattice generated by the parallelepiped. The resulting configuration is a lattice of dipoles consisting of one positive unit charge placed at $({y}^1_0,{y}^2_0,{y}^3_0)$ and one negative unit charge at $(a-{y}^1_0,{y}^2_0,{y}^3_0)$, together with all other dipoles generated from that one by translation by any number of lattice vectors. To fit to our setup, we set $a=b=c=1$ and $2\pi L =1$. One obtains eight charges (four dipoles) inside the unit $\mathbb{T}^3$. The total Green's function for this configuration reads \cite{Courant:1989aa}
\eq{\spl{\label{grsol}
H_{\vec{y}_0}(\vec{y})= \int_0^\infty \d t\Big(
&\left[
\theta_{3}\left( y^1_0+y^1,4\i \pi t \right)-\theta_{3}\left( y^1_0-y^1, 4\i \pi t \right)
\right]\times\\
&\left[
\theta_{3}\left( y^2_0+y^2,4\i \pi t \right)-\theta_{3}\left( y^2_0-y^2, 4\i \pi t \right)
\right]\times\\
&\left[
\theta_{3}\left( y^3_0+y^3,4\i \pi t \right)-\theta_{3}\left( y^3_0-y^3,4\i \pi t \right)
\right]
\Big)
~,}}
where the $\theta_3$-function is defined in \eqref{thetadef}.

Remarkably, the expression \eqref{grsol} can be obtained making use of the generalized Green's function $G$ \eqref{a3}, that we will discuss below. A positive/negative unit charge at $\vec{y}_0$ contributes $\pm G(\vec{y}-\vec{y}_0)$ to the total Green's function or warp factor $H$ \eqref{H}. In the configuration above there are four positive unit charges placed at $(y_0^1,y_0^2,y_0^3)$, $(1-y_0^1,1-y_0^2,y_0^3)$, $(1-y_0^1,y_0^2,1-y_0^3)$, $(y_0^1,1-y_0^2,1-y_0^3)$ and four negative unit charges at $(1-y_0^1,y_0^2,y_0^3)$, $(y_0^1,1-y_0^2,y_0^3)$, $(y_0^1,y_0^2,1-y_0^3)$, $(1-y_0^1,1-y_0^2,1-y_0^3)$. Summing up all eight contributions as in \eqref{H}, taking \eqref{a3} and \eqref{a5} into account, leads to \eqref{grsol}. We thus verify that in the $\mathbb{T}^3$ case, the generalized Green's function \eqref{a3} leads to the correct result established by Courant and Hilbert \cite{Courant:1989aa}.

\subsection{Generalized Green's functions}\label{sec:Green}

A first naive solution to \eqref{BIG} on a torus $\mathbb{T}^d$ is obtained using Fourier series, since the functions should be periodic. Using the coordinates introduced in \eqref{torusmetric}, we obtain
\beq
G(\vec{\sigma}) = -\frac{1}{(2\pi L)^{d-2}} \sum_{\vec{n}\in \mathbb{Z}^{d\, *}} \frac{e^{2\pi \i \vec{n}\cdot \vec{\sigma}}}{4\pi^2 \vec{n}^2} \ , \label{GFourier}
\eeq
where $\mathbb{Z}^{d\, *}$ is $\mathbb{Z}^{d}$ without $\vec{0}$. One may however express doubts on this expression, because the sum is not absolutely convergent for $d \geq 2$. Indeed, for $d=2$, one has
\beq
\sum_{(m,n)\in \mathbb{N}^{2\, *}} \frac{1}{m^2+n^2} \geq \sum_{(m,n)\in \mathbb{N}^{2\, *}} \frac{1}{(m+n)^2} = \sum_{k \in \mathbb{N}^{*}} \sum_{m+n=k} \frac{1}{k^2}  = \sum_{k \in \mathbb{N}^{*}}\frac{k+1}{k^2} \sim \sum_{k \in \mathbb{N}^{*}}\frac{1}{k} \ , \label{pbsum}
\eeq
and this gets generalized for $d>2$ using this and $n_1^2 + \dots + n_d^2 \leq n_1^2 + (n_2 + \dots + n_d)^2$. Given this issue, the following regularization was proposed in \cite{Shandera:2003gx, Kim:2018vgz}: one first uses
\beq
\frac{1}{4\pi^2 \vec{n}^2}=\int_0^\infty \d t ~\!e^{-4\pi^2\vec{n}^2 t}~,
\eeq
and further interchanges in \eqref{GFourier} the order of the integral and the sum. This last operation is a priori not allowed, since the series does not converge uniformly in the neighborhood of $t=0$ \cite{Courant:1989aa}, but this is precisely what brings the regularization. This manipulation leads to
\begin{empheq}[innerbox=\fbox]{align}
G(\vec{\sigma})
& =(2\pi L)^{2-d} \int_0^\infty \d t \Big(
1-\prod_{m=1}^d\theta_3(\sigma^m|4\pi \i t)
\Big) \label{a3}
\end{empheq}
with the theta function $\theta_3 = \theta_{00}$ given by
\eq{\label{thetadef}
\theta_3(\sigma|\tau)=\sum_{n \in \mathbb{Z}}
e^{2\pi \i(n\sigma +\frac{n^2}{2}\tau)} = 1 + 2 \sum_{n=1}^\infty q^{n^2} \cos(2\pi n \sigma ) \ ,\ {\rm with}\ q= e^{\i \pi \tau}
~.}
We loosely use the same symbol for this ``regularized'' generalized Green's function, and the above Fourier series expression; the distinction, if needed, should be clear from the context. One is free to add a constant to $G$, since only its derivatives enter \eqref{BIG}. Such a constant will not play a role in the following (in $H$ \eqref{H} in particular) where we will mostly consider $\sum Q_i = 0$, so we discard here this possibility.

While we are interested in using the proposal \eqref{a3} in the warp factor $H$ \eqref{H}, we may first check it in various manners. A first crucial test is that the $H$ obtained this way matches the rigorous derivation by Courant and Hilbert \cite{Courant:1989aa} for $d=3$, as explained at the end of section \ref{sec:threebrane}. It is easy to see that the sum of triple products of $\theta_3$ in \eqref{grsol} are reproduced by summing the proposed generalized Green's functions \eqref{a3} for $d=3$, the constant terms given by $1$ in the integral canceling out with the various signs of the charges. The derivation of Courant-Hilbert \cite{Courant:1989aa} discusses precisely the interchange of integral and sum, and conclude in a well-defined result: this gives us confidence in using the warp factor $H$ \eqref{H} with the proposal \eqref{a3} for any $d$.

\begin{figure}[H]
\begin{center}
\begin{subfigure}[H]{0.4\textwidth}
\includegraphics[width=\textwidth]{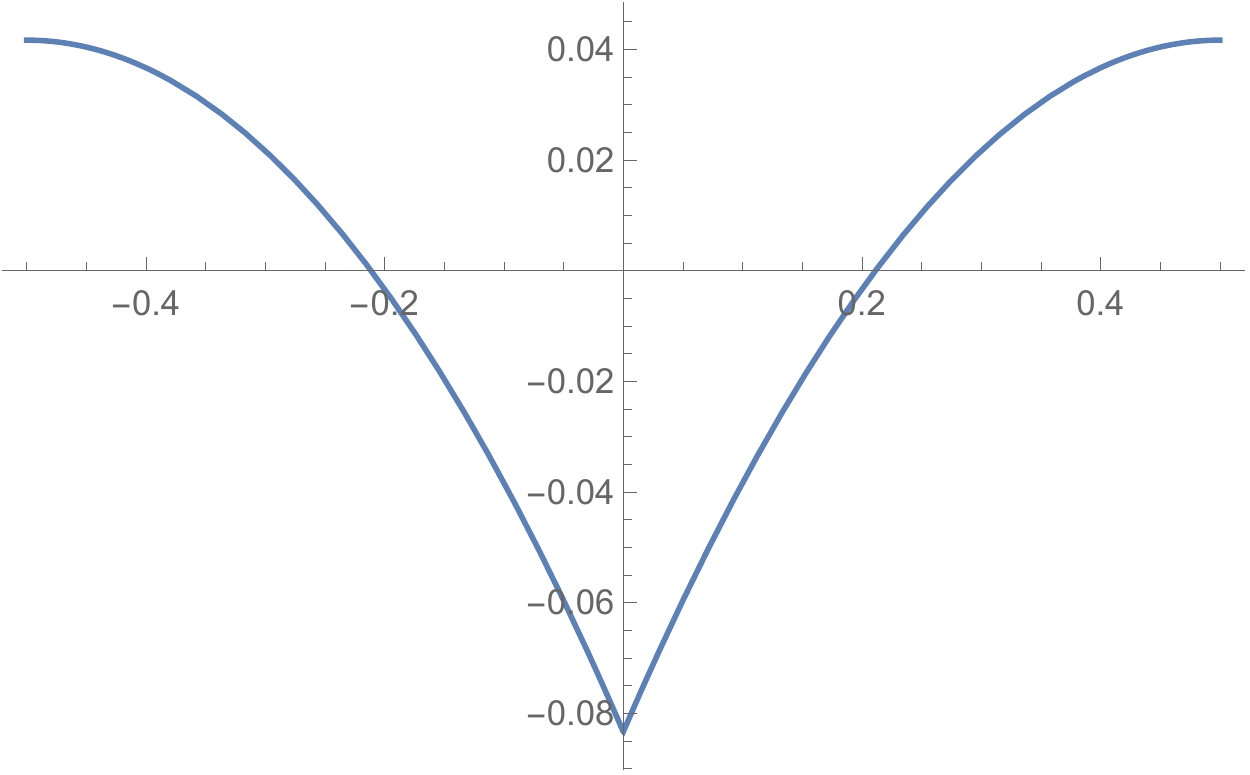}\caption{$d=1$}\label{fig:G1d}
\end{subfigure}
\qquad \qquad
\begin{subfigure}[H]{0.4\textwidth}
\includegraphics[width=\textwidth]{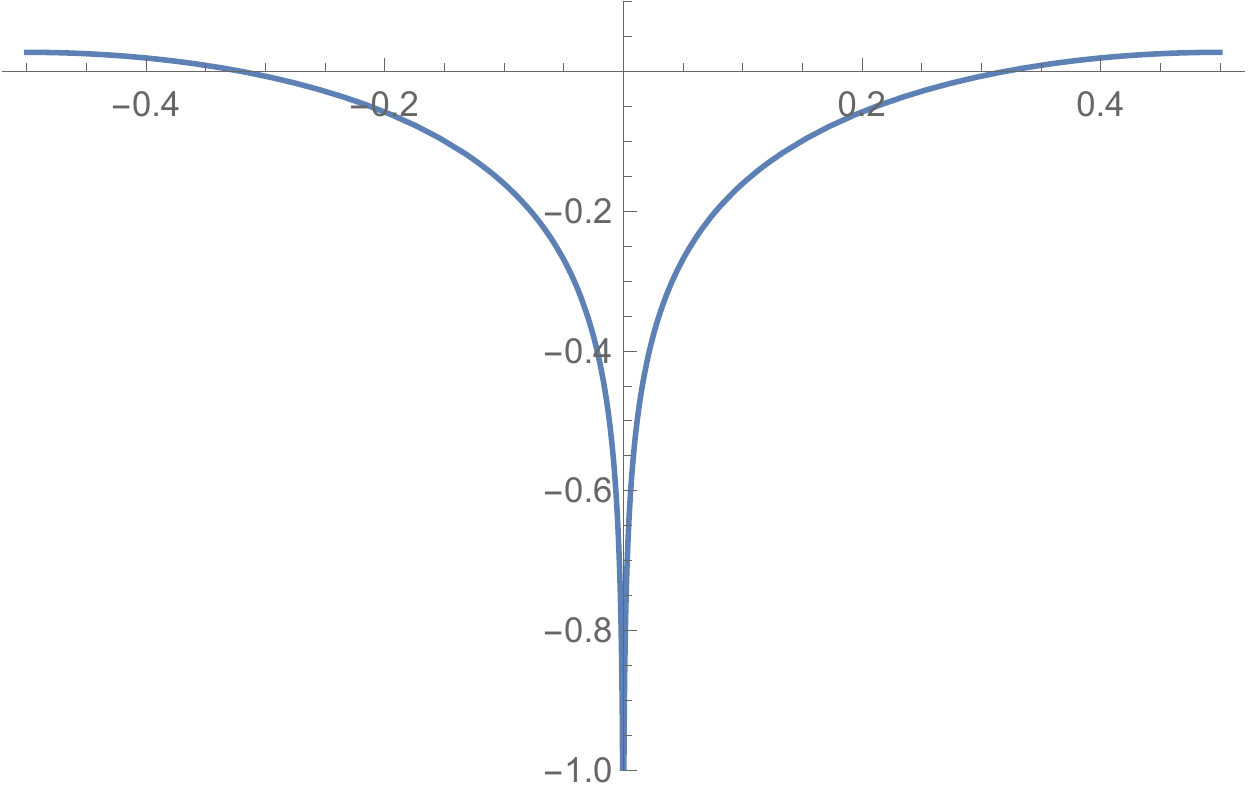}\caption{$d=2$}\label{fig:G2d}
\end{subfigure}
\begin{subfigure}[H]{0.4\textwidth}
\includegraphics[width=\textwidth]{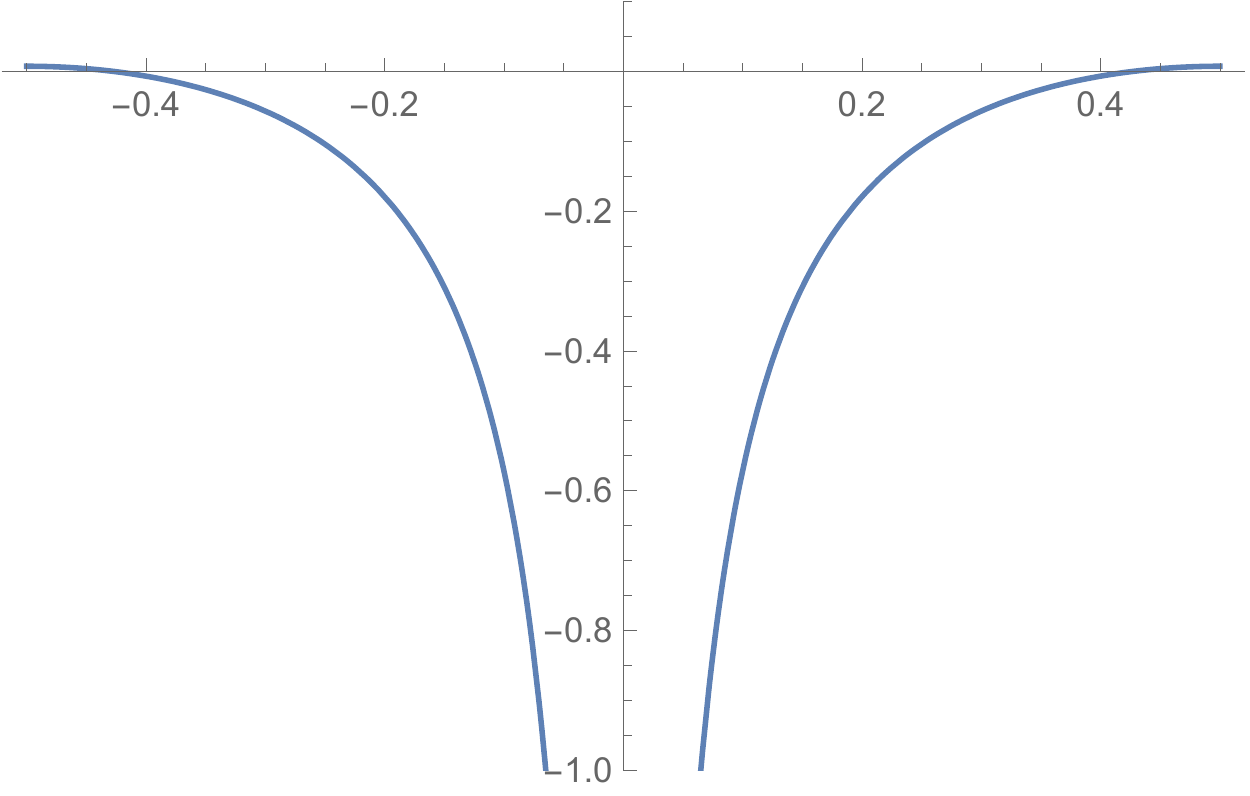}\caption{$d=3$}\label{fig:G3d}
\end{subfigure}
\qquad \qquad
\begin{subfigure}[H]{0.4\textwidth}
\includegraphics[width=\textwidth]{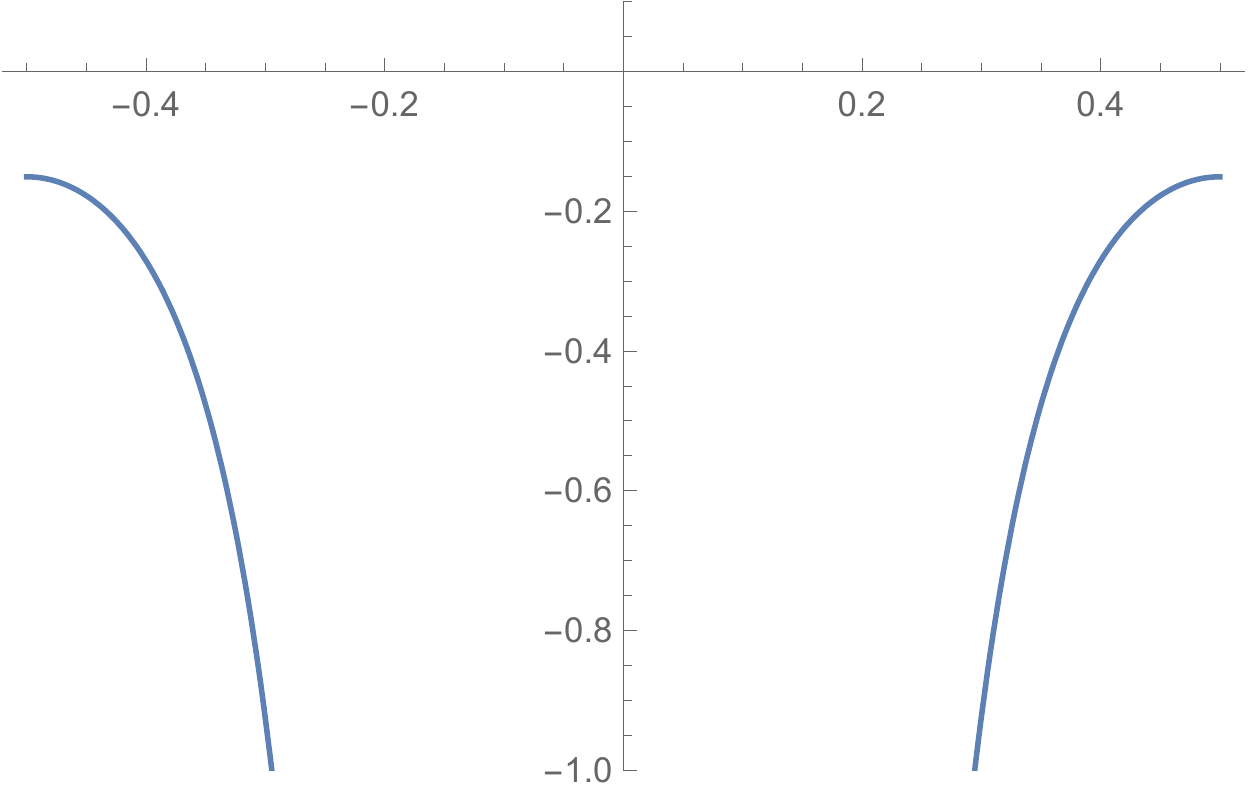}\caption{$d=6$}\label{fig:G6d}
\end{subfigure}
\caption{Green's function $(2\pi L)^{d-2}\ G(\vec{\sigma})$ given in \eqref{a3} for dimensions $d=1,2,3,6$, plotted along $\sigma^1 \in [-\tfrac{1}{2},\tfrac{1}{2}]$. The range of the frames are the same for $d=2,3,6$, allowing a direct comparison of the different behaviours close to the source at $\vec{\sigma}= \vec{0}$. More comments can be found in appendix \ref{ap:behaviour}.}\label{fig:G}
\end{center}
\end{figure}

Another test is the behaviour of the proposed generalized Green's function \eqref{a3} close to the $D_p$ or $O_p$ sources, for which one expects the same behaviour as in flat space. We show analytically in appendix \ref{ap:behaviour} that one precisely recovers this behaviour from the expression \eqref{a3}, modulo the introduction of some cut-off in the cases $d=1,2$. This study is illustrated in Figure \ref{fig:G} with plots of $G$ \eqref{a3} for various dimensions: the behaviour close to the source is apparent, as well as the periodicity of the function.

Finally, the generalized Green's function on a torus for $d=2$ is well-known in other contexts in the string literature, see e.g.~\cite{Polchinski:1998rr}. The expression is given in \eqref{G2d}, and is not obviously matching the proposed one \eqref{a3} for $d=2$. We compare in appendix \ref{ap:2d} the two expressions, and conclude that they are very likely to match, at least for $\tau_1=0$, as illustrated in Figure \ref{fig:G2dcompare}.

These various checks are successful, so we use the generalized Green's function $G$ \eqref{a3} on $\mathbb{T}^d$ in the warp factor $H$ \eqref{H}. There remains one missing ingredient in $H$, the constant $H_0$, that we now turn to.

\subsection{Normalisation of $H$: the constant piece}\label{sec:normH}

As pointed out in section \ref{sec:brane10}, the warp factor $H$ \eqref{H} includes a constant $H_0$ that is not fixed by the differential equation \eqref{BI}. It turns out to related to the normalisation of $H$ as we now see. We use the torus $\mathbb{T}^d$ with metric \eqref{torusmetric} as the unwarped transverse space, and compute an average of $H$ on it through the following integral
\beq
\int_{-\frac{1}{2}}^{\frac{1}{2}} \d^d \vec{\sigma}\ H = \sum_i Q_i\, \int_{-\frac{1}{2}}^{\frac{1}{2}} \d^d \vec{\sigma}\ G(\vec{\sigma} - \vec{\sigma}_i) + H_0 = \left( \sum_i Q_i \right)\times \int_{-\frac{1}{2}}^{\frac{1}{2}} \d^d \vec{\sigma}\ G(\vec{\sigma}) + H_0 \ , \label{intH}
\eeq
where one uses the periodicity of $G$ in $\sigma^m \sim \sigma^m+1$. The $d$-dimensional integral of $G$ is finite: at the source where $G$ diverges for $d\geq 2$, one verifies, thanks to the results of appendix \ref{ap:behaviour}, that the $d$-primitive (indefinite integral) without constant actually vanishes. Therefore we obtain
\beq
{\rm For}\ \sum_i Q_i = 0: \quad \int_{-\frac{1}{2}}^{\frac{1}{2}} \d^d \vec{\sigma}\ H = H_0 \ . \label{Haverage}
\eeq
For $ \sum_i Q_i = 0$, often considered in the following, the average of $H$ is simply the constant $H_0$: this gives more physical significance to that constant.

An important occurrence of this integral is the following.\footnote{One may also consider the 4d Planck mass $M_4$: using the $D=10$ string frame metric \eqref{10dmetricstringframe} and dilaton, one obtains
\beq
M_4^2= \frac{1}{2 \kappa_{10}^2} \int \d^6 y \sqrt{|g_6|} e^{-2\phi} H^{-\frac{1}{2}} = \frac{1}{(2\pi)^7 l_s^8 g_s^2} \int \d^6 y \sqrt{|\tilde{g}_6|}\ H =  \frac{2\pi}{(2\pi)^2 l_s^2 g_s}\ \frac{\int \d^{6-d} x \sqrt{|\tilde{g}_{6||}|}}{(2\pi l_s)^{6-d}}\ \frac{L^d}{l_s^{d}}\ \frac{1}{g_s} \int_{-\frac{1}{2}}^{\frac{1}{2}} \d^d \vec{\sigma}\ H \nn
\eeq
where the last $H^{-\frac{1}{2}}$ factor came from the 4d piece of the action. Interestingly, this expression involves the integral of $H$ itself. But it cannot be fixed in this manner since the integral also depends on the unwarped parallel space volume.} We recall from section \ref{sec:1mg} that the integral \eqref{norm} for $N=0$ should be finite and non-zero to have a normalisable zero-mode, and thus ensure the existence of a standard (massless) 4d gravitational wave. To compute this integral, we use the warp factor \eqref{warpmatch}, the Einstein frame metric \eqref{MmetricEinsteinD} with transverse torus \eqref{torusmetric} to get
\beq
\int \d^{D-4} y\, \sqrt{\hat{g}}\, e^{2A} = \int \d^{D-4-d} x\, \sqrt{|\tilde{g}_{||}|}\, \times (2\pi L)^d \int_{-\frac{1}{2}}^{\frac{1}{2}} \d^d \vec{\sigma}\ H \ ,
\eeq
where $\sqrt{|\tilde{g}_{||}|}= 1$ in the toroidal case. Remarkably, all warp factors conspire to leave us with the integral of $H$. With a compact parallel internal space and $ \sum_i Q_i = 0$, we get as wished a finite integral, and deduce the following requirement for a non-vanishing integral
\beq
H_0 \neq 0 \ .
\eeq

There is another important reason to have a non-zero constant in $H$. The warp factor enters the metric, through powers of it and their inverse, so $H$ should not vanish, to avoid unwanted singularities and possible signature changes. This is reflected e.g.~in \eqref{warpmatch} where $H^{\frac{D-p-3}{D-2}}=e^{-2A}>0$. Adding a constant in $H$ and modifying this way its normalisation should help getting it of definite sign, as we will see. In standard non-compact $p$-brane solutions, the constant $H_0$ is fixed by asymptotics, while the sign of $H$ is guaranteed beyond a horizon. But there is no notion of asymptotics on a compact space. In the following, we then propose a natural prescription to fix the constant $H_0$, sticking to the idea of having $H>0$. For future convenience, we introduce the dimension-dependent constant $h_d$ as follows
\beq
H = \sum_i Q_i\, G(\vec{y} - \vec{y}_i) + H_0\ , \quad H_0 = g_s\, h_d \ . \label{Hwithconstant}
\eeq
To fix $h_d$, we study the sign of $H$ and start with $d\geq 2$.

\subsubsection*{Fixing the constant for $d\geq 2$}

The study of appendix \ref{ap:behaviour} shows that the generalized Green's function $G$ becomes divergent close to its source, say at $|\vec{\sigma}|=0$. Close to this point, $H$ is dominated by the corresponding $G$'s, which are negative, as can be seen in \eqref{behav3d} and \eqref{behav2d} or in Figure \ref{fig:G}. The $O_p$ charge being positive, such a source then leads to a risk that $H$ becomes negative in this region, contrary to the case of a $D_p$. To avoid this, we propose to require that $H$ remains positive at a distance to the source bigger than $l_s$, i.e.~$L|\vec{\sigma}| > l_s$; beyond this point, there is no reason to trust the supergravity solution over string effects anyway.\footnote{We use here the unwarped metric \eqref{torusmetric} for the notion of distance (to the source) and its comparison to $l_s$. A proper distance should however include the warp factor itself. For simplicity we stick here to this prescription.} This idea amounts to setting the size of a horizon close to an $O_p$. So we fix the constant formally as follows
\bea
H_0 = g_s\, h_d = & - {\rm Min}_{\vec{y}} \, \left\{\sum_i Q_i\, G(\vec{y} - \vec{y}_i) \right\}\ \ {\rm s.t.}\ \tfrac{y^m}{2\pi L} \in [-\tfrac{1}{2},\tfrac{1}{2}]\ {\rm and}\ \forall j \ |\vec{y} - \vec{y}_j|\geq 2\pi l_s  \label{constantpresc} \\
= & - {\rm Min}_j \, \left\{\sum_i Q_i\, G(\vec{y} - \vec{y}_i)|_{|\vec{y} - \vec{y}_j|=2\pi l_s} \right\} \ . \nn
\eea
As explained for $d\geq 2$, this is roughly given only by the orientifold contribution at one $\vec{y}_j$. Using \eqref{QD}, \eqref{behav3d} and \eqref{behav2d}, this becomes
\eq{\label{constant}\spl{
h_{d\geq 3} \simeq & - \frac{Q_{O_p}}{g_s}\, G(\vec{\sigma})|_{L|\vec{\sigma}|=l_s} = 2^{2-d} \pi^{-\frac{d}{2}}\ \Gamma\left(\tfrac{d-2}{2} \right) \ ,\\
{\rm i.e.} \ \ h_{3} \simeq &\, \frac{1}{2\pi} \ ,\ h_{4} \simeq \frac{1}{(2\pi)^2} \ , \ h_{5} \simeq \frac{1}{2^4\pi^2} \ , \ h_{6} \simeq \frac{1}{2^4\pi^3} \ ,\\
{\rm and} \ \ h_{d=2} \simeq &\, \frac{2}{\pi} \ln \left(\frac{L}{l_s}\right) \ .
}}
To get the estimates \eqref{constant}, we first considered in \eqref{constantpresc} only an orientifold contribution, at $l_s$ distance from it, meaning that we assumed other sources to be at a much larger distance. Since $G$ is negligible away from its source compared to the divergence (see e.g.~Figure \ref{fig:G}), the single $O_p$ contribution would dominate. In practice, the other sources are placed at distances being fractions of $L$, so this assumption requires  $L\gg l_s$. Secondly, we used the behaviours \eqref{behav3d} and \eqref{behav2d} of $G$ close to the source to get estimates: this requires again $L\gg l_s$ since we consider $G(\vec{\sigma})|_{|\vec{\sigma}|=l_s/L}$. The approximation $L\gg l_s$ can be physically well-justified as we will discuss around \eqref{condphysrel}: in that case, \eqref{constant} are then good estimates of \eqref{constantpresc}, allowing to get $H>0$ as desired.

\begin{figure}[H]
\begin{center}
\begin{subfigure}[H]{0.4\textwidth}
\includegraphics[width=\textwidth]{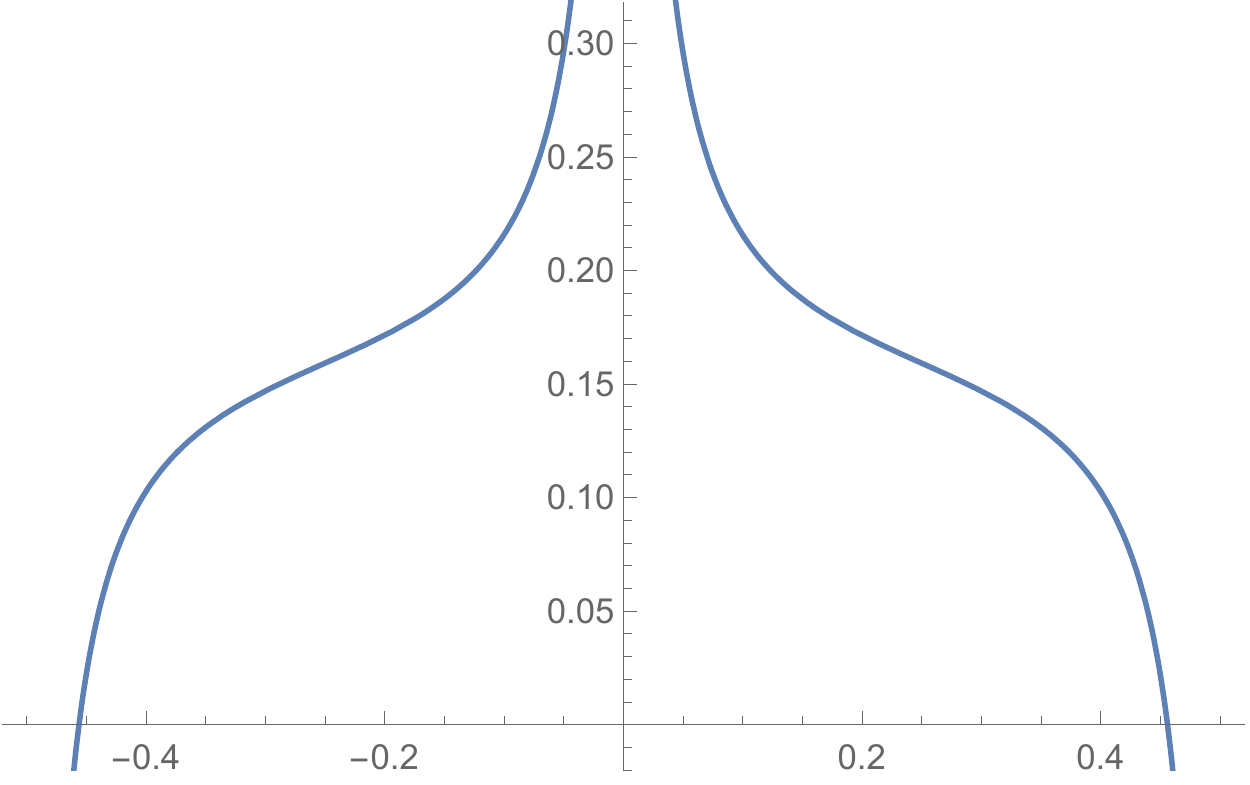}\caption{$\tfrac{L}{l_s}=20$}\label{fig:H20}
\end{subfigure}
\qquad \qquad
\begin{subfigure}[H]{0.4\textwidth}
\includegraphics[width=\textwidth]{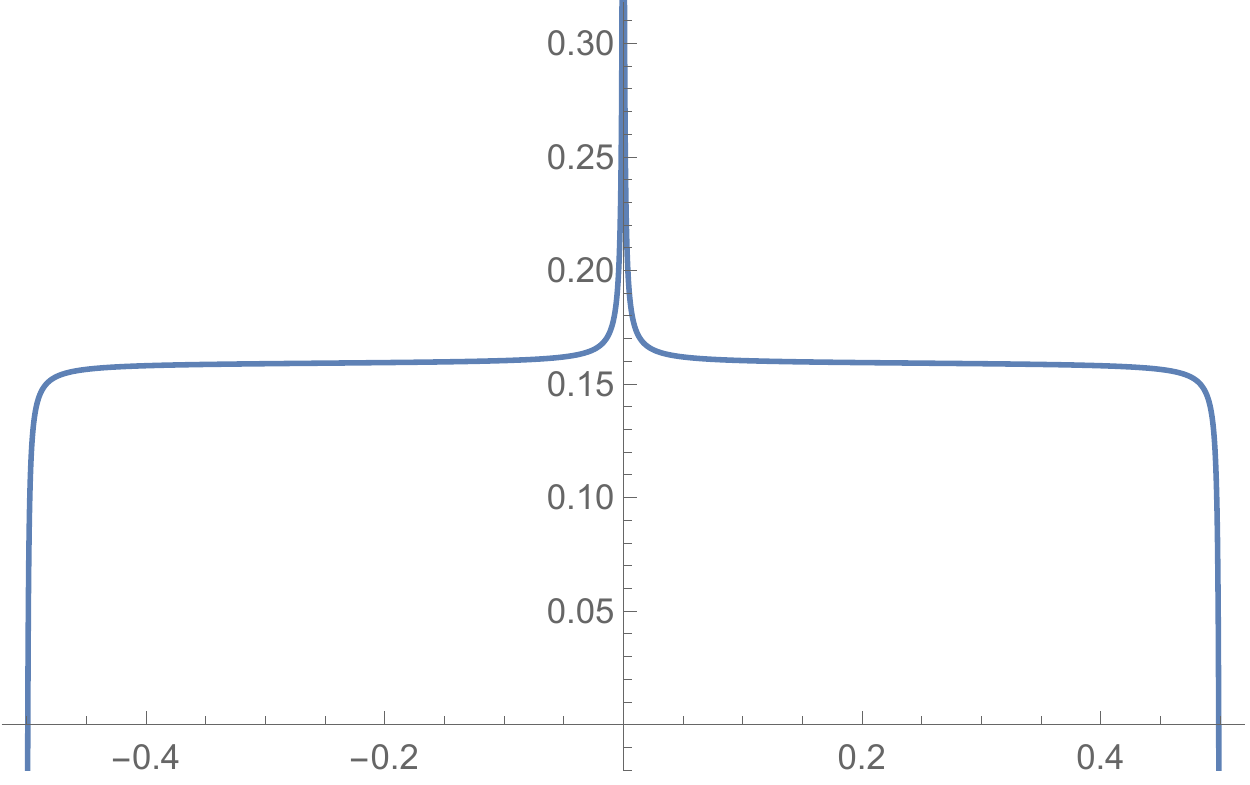}\caption{$\tfrac{L}{l_s}=1000$}\label{fig:H1000}
\end{subfigure}
\caption{Warp factor $H \times \tfrac{1}{g_s}$ given in \eqref{Hex} for one charge $Q_{O_6}$ at $\sigma^1=\pm\tfrac{1}{2}$ and the opposite charge at $\sigma^1=0$, plotted along $\sigma^1 \in [-\tfrac{1}{2},\tfrac{1}{2}]$ for two different values of $\tfrac{L}{l_s}$. We verify that $H>0$ until an approximate distance $\Delta \sigma^1 =\tfrac{l_s}{L}$ to the charge $Q_{O_6}$.}\label{fig:H3d}
\end{center}
\end{figure}

To illustrate the above, we consider a dipole toy model with $D=10$ and $p=6$, i.e.~$d=3$, with one object of charge $Q_{O_6}$ at $\vec{\sigma}_O=(\pm\tfrac{1}{2},0,0)$ and one of opposite charge at $\vec{\sigma}=(0,0,0)$. We obtain
\beq
\frac{H}{g_s} = 2 \frac{l_s}{L}\, \left(2\pi L \, G(\vec{\sigma} - \vec{\sigma}_O) - 2\pi L \, G(\vec{\sigma}) \right) +  \frac{1}{2\pi} \ . \label{Hex}
\eeq
This warp factor is depicted in Figure \ref{fig:H3d} for two values of the ratio $\tfrac{l_s}{L}$. The domination of the $O_6$ contribution close to $\vec{\sigma}_O$ and the appropriate sign of $H$ are well illustrated there.

\subsubsection*{Fixing the constant for $d=1$}

The situation is different for $d=1$: the corresponding generalized Green's function $G$ is depicted in Figure \ref{fig:G1d}. One can see that $G$ does not diverge at the source at $\sigma=0$, as also noticed through its behaviour obtained in \eqref{behav1d}. Even though $G(0)$ remains the minimum, it does not differ much from the value away from the source: we computed around \eqref{behav1d} that $G(\pm \frac{1}{2})= -\frac{1}{2} G(0) = \frac{2\pi L}{24} $. Therefore, the contribution of other sources to the total $H$ is not negligible anymore. To ensure $H>0$, we then simply propose the prescription
\beq
g_s\, h_1 = - {\rm Min}_{\vec{y}}\, \left\{\sum_i Q_i\, G(\vec{y} - \vec{y}_i) \right\} \ , \label{constantpresc1d}
\eeq
for which there is in general no easy estimate.

\begin{figure}[H]
\begin{center}
\begin{subfigure}[H]{0.4\textwidth}
\includegraphics[width=\textwidth]{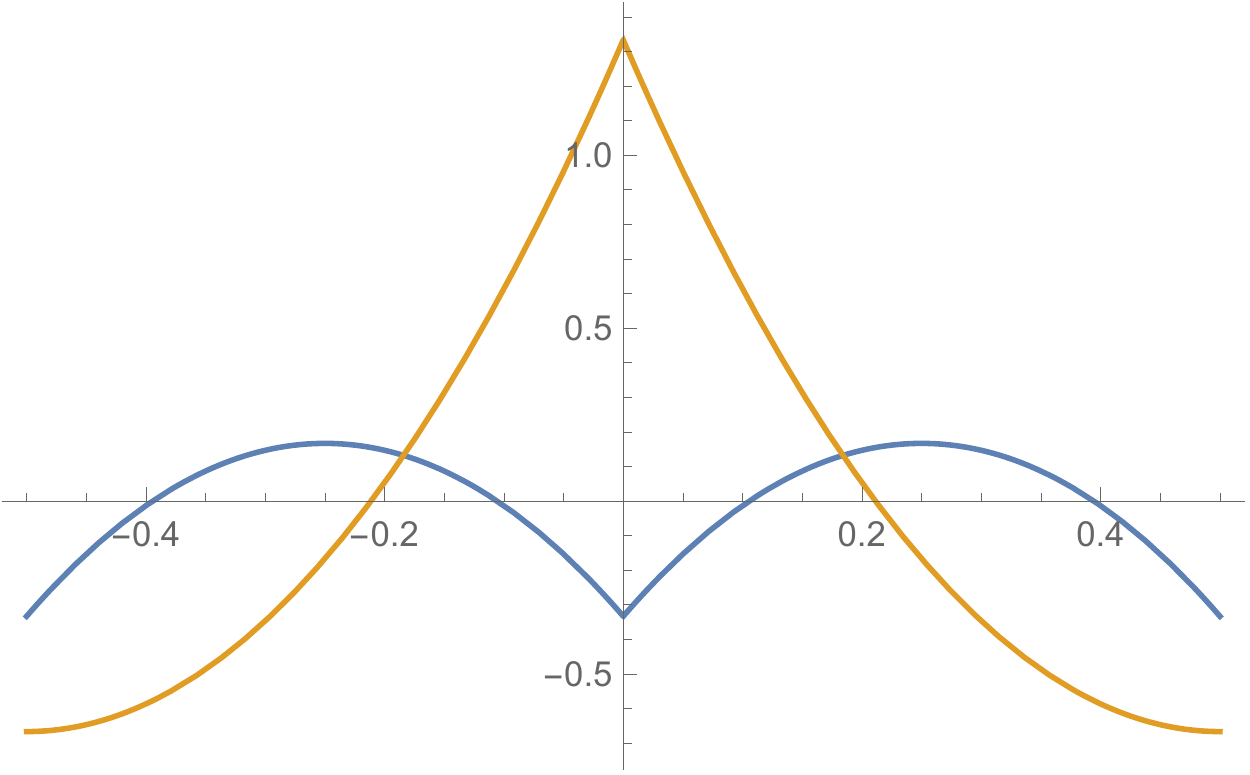}\caption{}\label{fig:H1dsep}
\end{subfigure}
\qquad \qquad
\begin{subfigure}[H]{0.4\textwidth}
\includegraphics[width=\textwidth]{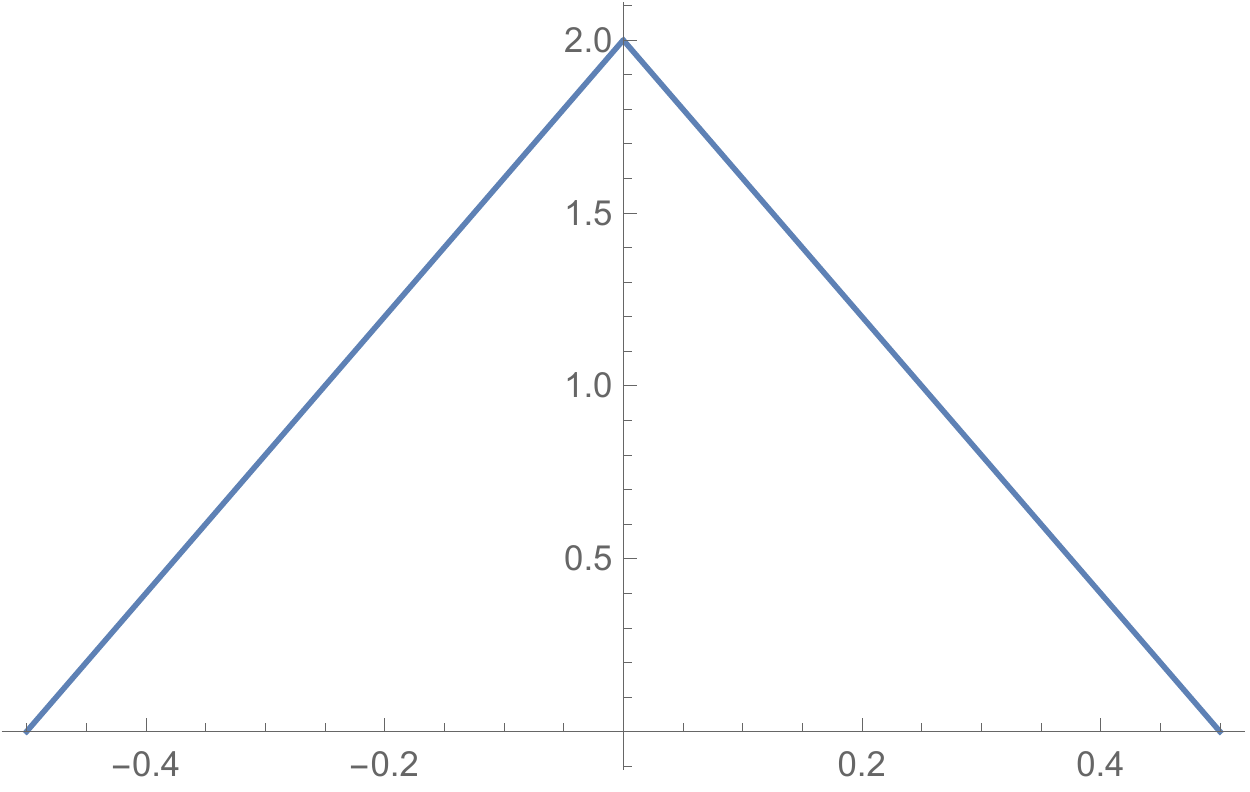}\caption{}\label{fig:H1dall}
\end{subfigure}
\caption{Warp factor $H \times \frac{l_s}{g_s L}$ given in \eqref{H1d} for one $O_8$ at $\sigma=0$, one at $\sigma=\pm\tfrac{1}{2}$ and $16$ $D_8$ at $\sigma=0$. The separate source contributions to $H$ are shown in Figure \ref{fig:H1dsep}: the bottom blue curve is that of the $O_8$, while the top orange curve is that of the $D_8$. The overall warp factor \eqref{H1d}, including the constant, appears in Figure \ref{fig:H1dall}.}\label{fig:H1d}
\end{center}
\end{figure}

Let us compute this constant in the following relevant example. We consider the case of orientifolds in $D=10$ with $d=1$, i.e.~$p=8$: these $O_8$ are placed at $\sigma=0, \pm \frac{1}{2}$. To compensate their charge $2 Q_{O_8}$ (studying a $\sum_i Q_i=0$ case), we need 16 $D_8$: we place them all at $\sigma= 0$. The minimum of the function is thus at $\sigma=\pm \frac{1}{2}$ (see e.g.~the different source contributions in Figure \ref{fig:H1dsep}), so we get from \eqref{constantpresc1d}
\bea
g_s\, h_1 = & \, - \left(16 Q_{D_8} G(\pm \frac{1}{2}) + Q_{O_8} G(\pm \frac{1}{2}) + Q_{O_8} G(0) \right) = - Q_{O_8} \left(G(0) - G(\pm \frac{1}{2})\right) \ ,\nn \\
{\rm i.e.}\  h_1 = &\, \frac{L}{l_s} \ .  \label{constant1d}
\eea
In short, we obtain in that example the following warp factor
\beq
H \times \frac{l_s}{g_s L} =  8  \left( (2\pi L)^{-1}\, G(\sigma - \frac{1}{2}) - (2\pi L)^{-1}\, G(\sigma) \right) + 1 \ . \label{H1d}
\eeq
We note from \eqref{H1d} that this source configuration eventually amounts to a dipole. This warp factor is depicted in Figure \ref{fig:H1dall}, from which we verify $H>0$.
\noindent As a curiosity, one can prove that the warp factor \eqref{H1d} is equal to the triangle function $f(\sigma)= -4 |\sigma| + 2$ as seen on Figure \ref{fig:H1dall}. For $d=1$, we can actually use the Fourier series \eqref{GFourier} which does not suffer from the problem \eqref{pbsum}. We compute from it and \eqref{H1d} the expression
\beq
H \times \frac{l_s}{g_s L} = 1 + \frac{4}{\pi^2}\, \sum_{n \in 2 \mathbb{Z}+1}\, \frac{e^{2\pi \i n \sigma}}{n^2}  \ . \label{H1dFourier}
\eeq
This can easily be verified to be the Fourier series of the periodic $f(\sigma)$ by computing its Fourier coefficients.

\section{Kaluza--Klein spectrum of gravitational waves}\label{sec:spectrumgal}

Four-dimensional gravitational waves propagating on a 4d warped Minkowski background with toroidal extra dimensions admit a Kaluza--Klein mass spectrum $M_N$ determined from equation \eqref{eigeneq}. We rewrite it here in terms of coordinates $\sigma^m$ on the transverse torus, introduced in \eqref{torusmetric}, and get
\eq{\label{intt7}
-\delta^{mn} \frac{\partial}{\partial\sigma^m}\frac{\partial}{\partial\sigma^n} \psi_N=H\, (2\pi L\, M_N)^2\,  \psi_N
~.}
Solutions $\psi_N$ to this equation are in one-to-one correspondence with 4d Kaluza--Klein gravitational waves of mass $M_N$ as discussed in section \ref{sec:1mg}. Having determined in section \ref{sec:Greengal} the warp factor $H$ on a transverse torus $\mathbb{T}^d$, in terms of generalized Green's functions and a constant $H_0$ as in \eqref{Hwithconstant}, we now proceed to determine the spectrum from \eqref{intt7}.

We first need to be more specific about the $D$-dimensional background to fix completely $H$: the formalism developed so far allows e.g.~to have the warp factor generated from dipoles of charges $Q_{D_p}$ and $Q_{O_p}$ or from stringy $D$-branes and orientifolds; we restrict in this section to the latter and give an explicit example in section \ref{sec:p=3}. We then make a first analysis of the spectrum in section \ref{sec:KKscale}, discussing the standard Kaluza--Klein spectrum and deviation from it, and finally provide in section \ref{sec:num} a numerical evaluation of the spectrum.

\subsection{Stringy background: a type IIB example}\label{sec:p=3}

Restricting to a stringy background specifies that the charged objects are stringy $D$-branes and orientifolds. This has one important consequence (beyond fixing $D=10$ and the value of charges): the number and placement of the orientifolds is fixed. For $\sum_i Q_i=0$, this in turn fixes the number of $D$-branes. We illustrate this situation here with an example for $d=6$, connecting to backgrounds presented in section \ref{sec:brane10}; another example was given after \eqref{constantpresc1d} with $d=1$.

We focus on the case of $p=3$ branes and orientifolds in $D=10$, as described in type IIB supergravity. We consider a well-known solution of the form $\mathbb{R}^{1,3}\times \mathbb{T}^6$ (warped product) with 16 $D_3$'s and 64 $O_3$'s \cite{Polchinski:1998rr, Johnson:1998pc}, which has been used in different contexts in e.g.~\cite{Verlinde:1999fy, Giddings:2001yu}, and is a particular solution in the family of \cite{Andriot:2016ufg} presented in section \ref{sec:brane10}. Orientifolding acts on the transverse space $\mathbb{T}^6$ as a parity-reversing $\mathbb{Z}_2$ involution, $\vec{y}\rightarrow-\vec{y}$. There are as many orientifolds as there are fixed points under the involution, namely $2^6$. Therefore the 16 $D_3$-branes ensure that the total charge vanishes as it should, since the $O_3$ tension and charge is $-\tfrac{1}{4}$ that of the $D_3$. The location of the orientifolds is at all points $\vec{y}_{i}=({y}_{i}^1,\dots, {y}_{i}^6)$, $i=1 \dots 64$, where ${y}_{i}^m=0$ or ${y}_{i}^m=\pi L$. The location of the $D_3$-branes on the other hand is arbitrary, and we place them for simplicity all at the origin $\vec{y} = \vec{0}$.\footnote{One may also consider ``smearing'' them with some density $\rho(\vec{y})$ in the transverse space subject to the constraint $\int_{\mathbb{T}^6}\d^6 y ~\!\rho(\vec{y})=16,\ \rho(\vec{y})=\rho(-\vec{y})$.} Moreover, the metric, the dilaton and the five-form field-strength are all invariant under the orientifold involution, i.e.~they are parity-even. The $D=10$ metric in string frame is given in \eqref{10dmetricstringframe} with a flat toroidal metric on the internal space, and the dilaton is $e^{\phi}=g_s$. The solution is completely determined by the function $H(\vec{y})$: it is given from \eqref{Hwithconstant} and \eqref{52} as follows
\eq{H(\vec{y}) = 16\, Q_{D_3} \, G(\vec{y}) - \frac{1}{4}\, Q_{D_3} \sum_{i} G(\vec{y} - \vec{y}_{i}) + g_s\, h_6 \ ,\quad  Q_{D_3} = - 4 Q_{O_3} = - (2\pi l_s)^{4} g_s\ , \label{HIIB}}
where $\vec{y}_{i}$ parameterizes the $2^6$ orientifold locations as discussed above; we can also introduce $\vec{\sigma}_{i}=\vec{y}_{i}/(2\pi L)$. The generalized Green's function on the right-hand side is that of \eqref{a3} with $d=6$. Using this solution as the gravitational wave background, its Einstein frame version (see \eqref{metricEinstein10}) leads as shown in section \ref{sec:braneD} to the eigenfunction equation \eqref{eigeneq}, rewritten in \eqref{intt7}, that determines the gravitational wave mass spectrum $M_N$. We give in the next section a general analysis of this spectrum and will evaluate it on analogous backgrounds with $d=1,2,3$ in section \ref{sec:num}.

\subsection{Standard Kaluza--Klein spectrum, deviation, and physical regime}\label{sec:KKscale}

We recall from \eqref{metricEinstein} and \eqref{torusmetric} that the Einstein frame metric is
\beq
\d s_E^2 = H^{-\frac{D-p-3}{D-2}} \left( \eta_{ij}\d x^i\d x^j+ H\, L^2 4 \pi^2 \delta_{mn}  \d \sigma^m \d \sigma^n \right) \ ,
\eeq
which indicates that the physically relevant length or toroidal radius is given by $\sqrt{H} L$; the same can be seen from the string frame metric. In absence of sources, i.e.~for $Q_i=0$, one has $H=H_0$, while in presence of sources, we recall from \eqref{Haverage} the average $\int_{-\frac{1}{2}}^{\frac{1}{2}} \d^d \vec{\sigma}\ H = H_0$. Therefore, the standard Kaluza--Klein spectrum to be considered here, i.e.~the one obtained in absence of sources, but also the one generated by the constant piece of the warp factor, or its average, is the spectrum obtained on a torus of radius $\sqrt{H_0} L$, namely
\beq
\mbox{Standard KK spectrum:}\quad \mathring{M}_{N}^2 = \frac{N^2}{H_0 L^2 } \ ,\quad N=|\vec{m}|\ ,\ \vec{m}\in \mathbb{Z}^{d} \ . \label{KKstandard}
\eeq
To measure the effect of a (non-constant) warp factor on the Kaluza--Klein spectrum, we need to compare it to the standard spectrum \eqref{KKstandard}. A deviation from the latter would only be due to the non-constant piece of the warp factor, or equivalently, to the sources. In other words, looking at $H$ in \eqref{Hwithconstant}, the standard spectrum \eqref{KKstandard} will be reproduced if $H_0 = h_d g_s$ dominates over the $Q_i G$. We can give a rough condition on the parameters in the model for this to happen: $(2 \pi L)^{d-2} G$ given in \eqref{a3} gives pure numbers, independent of parameters, and from \eqref{QD} one has $Q_i \sim (2 \pi l_s)^{d-2} g_s$. So we obtain the following condition
\beq
h_d \gg \left( \frac{l_s}{L}\right)^{d-2} \ \Rightarrow \ \mbox{standard KK spectrum} \ . \label{condstandardKK}
\eeq
The above intuition is verified comparing Figure \ref{fig:H20} and \ref{fig:H1000}: for a high $L/l_s$ ratio, the warp factor is much closer to a constant than for a smaller ratio, for which the non-constant contribution clearly appears. By introducing the tools needed to evaluate the spectrum, we are now going to rederive the condition \eqref{condstandardKK}, and give further comments.\\

The Kaluza--Klein spectrum is determined by equation \eqref{intt7}, for any $D$ and $d=D-p-1$. Perhaps the simplest approach to solving this equation is to use Fourier series on the transverse torus $\mathbb{T}^d$. Let us first expand
\eq{\label{fexpc}
\psi_N(\vec{\sigma})
= \sum_{\vec{m}\in \mathbb{Z}^{d}}  {e^{2\pi \i \vec{m}\cdot \vec{\sigma}}} c_{\vec{m}}
~.}
Using \eqref{Hwithconstant} for $H$ and substituting \eqref{GFourier} for the generalized Green's functions,\footnote{One may worry about using the Fourier expansion for the generalized Green's function, given that we argued in favor of a regularization in section \ref{sec:Green}. In the present approach, the regularization will be effectively provided by cutting off the infinite sum of Fourier modes of $\psi_N$. This is justified by the fact that $\psi_N$ is assumed normalisable, as discussed around \eqref{norm}.} together with \eqref{fexpc} into equation \eqref{intt7}, we obtain an infinite system of equations,
\eq{\label{infsys}
\forall \vec{m}\in \mathbb{Z}^{d} \ ,\quad \left(\frac{\vec{m}^2}{L^2 M_N^2} - H_0 \right)c_{\vec{m}}+\frac{1}{(2\pi L )^{d-2}}
 \sum_{\vec{n}\in \mathbb{Z}^{d\, *}} \frac{\sum_i Q_i e^{-2\pi \i \vec{n}\cdot \vec{\sigma_i}}}{ (2\pi\vec{n})^2}c_{\vec{m}-\vec{n}} = 0 \ ,}
where we consider $M_N\neq 0$. We introduce for each source located at $\vec{\sigma_i}$ a number $q_i=Q_i/Q_{D_p}$, i.e.~$q_i=1$ for a $D_p$ and $-2^{4-d}$ for an $O_p$; we also use the charges given in \eqref{QD}. We then rewrite the system of equations \eqref{infsys} as $\ocal_{\vec{m}}{}^{\vec{p}} c_{\vec{p}}=0$, with a matrix operator $\ocal$ given by
\beq
\ocal_{\vec{m}}{}^{\vec{p}} = \left(\frac{\vec{m}^2}{L^2 M_N^2} - g_s h_d \right)\delta_{\vec{m}}^{\vec{p}} -g_s \left(\frac{l_s}{L}\right)^{d-2}
 \sum_{\vec{n}\in \mathbb{Z}^{d\, *}} \frac{\sum_i q_i e^{-2\pi \i \vec{n}\cdot \vec{\sigma_i}}}{ (2\pi\vec{n})^2}\delta_{\vec{m}-\vec{n}}^{\vec{p}} \ . \label{operator}
\eeq
Since $\psi_N$ is assumed normalisable (see \eqref{norm}), its Fourier expansion converges, in particular $c_{\vec{n}}\rightarrow 0$ for $|\vec{n}|\rightarrow \infty$. We may then truncate to $c_{\vec{m}}=0$ for $|\vec{m}|> m_0$, where $m_0$ is some positive number. The error
thus committed becomes arbitrarily small as $m_0\rightarrow\infty$. As a result of this truncation, the infinite system of equations $\ocal_{\vec{m}}{}^{\vec{p}} c_{\vec{p}}=0$ reduces to a finite set of homogeneous equations for the coefficients $c_{\vec{p}}$ with $|\vec{p}|\leq m_0$, and the operator $\ocal$ can be taken as a finite size matrix. This system of equations admits non-trivial solutions only when imposing $\text{det}(\ocal)=0$, which in turn would select discrete values of $M_N^2$, corresponding to the allowed Kaluza--Klein masses. This will serve as the numerical method to evaluate the spectrum in section \ref{sec:num}. Before doing so, let us take a closer look at this matrix.

We want to compare the various coefficients of $\ocal$, especially the off-diagonal ones due to the non-constant piece of $H$ and the contribution $H_0$ to the diagonal ones. For each $\vec{n}\in \mathbb{Z}^{d\, *}$, one has, using $\sum_i Q_i =0$,
\beq
\left| \frac{\sum_i q_i e^{-2\pi \i \vec{n}\cdot \vec{\sigma_i}}}{ (2\pi\vec{n})^2} \right| \leq \frac{\sum_i |q_i| }{ (2\pi)^2} =  \frac{2 \sum_{O_p} |Q_{O_p}| }{(2\pi)^2 |Q_{D_p}| } =  \frac{2^3}{\pi^2} \ .
\eeq
The last equality is obtained by considering that the charges $Q_{O_p}$ are placed at the $2^{d}$ fixed points, as for instance in the string theory setting described in section \ref{sec:p=3}. We then get the comparison of coefficients minimized as follows
\beq
g_s h_d \times \left|g_s \left(\frac{l_s}{L}\right)^{d-2}  \frac{\sum_i q_i e^{-2\pi \i \vec{n}\cdot \vec{\sigma_i}}}{ (2\pi\vec{n})^2} \right|^{-1} \geq   \left(\frac{L}{l_s}\right)^{d-2} h_d\, \frac{\pi^2} {2^3} \ . \label{ratio}
\eeq
Therefore, if $\left(L/l_s\right)^{d-2} h_d \gg 1$, the contribution of $H_0$ to the diagonal terms dominates the off-diagonal coefficients. In other words, the off-diagonal coefficients of $\ocal$ in \eqref{operator} can in that case be neglected to get the spectrum:
\beq
\left(\frac{L}{l_s}\right)^{d-2} h_d \gg 1 \ \Rightarrow\  \ocal_{\vec{m}}{}^{\vec{p}} \simeq \left(\frac{\vec{m}^2}{L^2 M_N^2} - g_s h_d \right)\delta_{\vec{m}}^{\vec{p}} \ . \label{condapproxO}
\eeq
It is then easy to read the spectrum with $\text{det}(\ocal)=0$
\beq\label{410}
M_{N}^2 \simeq \frac{N^2}{H_0 L^2 }  \ ,
\eeq
with $N=|\vec{m}|$ labelling the Kaluza--Klein masses. This is precisely the standard Kaluza--Klein spectrum $\mathring{M}_N$ \eqref{KKstandard}, and the condition to obtain it, given in the left-hand side of \eqref{condapproxO}, matches exactly the one obtained in \eqref{condstandardKK} by a rough estimation. This analysis implies that obtaining a deviation from the standard spectrum, which would be physically more interesting, requires the contrary to \eqref{condstandardKK}, namely
\begin{empheq}[innerbox=\fbox]{align}
\mbox{Deviation from standard KK spectrum} \ \Rightarrow \ h_d \lesssim \left( \frac{l_s}{L}\right)^{d-2} \label{conddevstandardKK}
\end{empheq}

Before commenting further on that condition, let us discuss another one. It is a standard requirement, for the classical solutions to be trusted, to ask for the fundamental length $l_s$ to be much smaller than the typical, or average, size of the extra dimensions: this can be written here as $\sqrt{H_0} L \gg l_s$, meaning
\begin{empheq}[innerbox=\fbox]{align}
\mbox{Physically relevant regime:}\quad h_d \gg \frac{1}{g_s} \left( \frac{l_s}{L}\right)^2  \label{condphysrel}
\end{empheq}
The two conditions just derived, and conclusions to be drawn, might be disputed away from a string-theory-like context. For instance, in a more phenomenological model, the charges may not obey the definition \eqref{QD} used, or the length entering those may not be understood as a fundamental length. Also, we use in the following a stringy interpretation to require $g_s < 1$, but this might not apply in a different context. With these limitations in mind, let us now combine the two above conditions, to see whether we have a chance to get a deviation from the standard Kaluza--Klein spectrum while remaining in a physically relevant regime. We deduce combining \eqref{conddevstandardKK} and \eqref{condphysrel}
\begin{empheq}[innerbox=\fbox]{align}
\mbox{Non-standard spectrum in a physically relevant regime}  \Rightarrow \quad \left( \frac{l_s}{L}\right)^{d-4} \gg 1
\end{empheq}

The first conclusion is that this will not be reached for $d=4$. For $d\geq 5$, we obtain the requirement $l_s/L \gg 1$, which from \eqref{condphysrel} gives $h_d \gg 1$. It is difficult to know whether this situation can agree with the prescription \eqref{constantpresc} for $h_d$ proposed in section \ref{sec:normH}. There we only evaluated $h_d$ to the explicit values \eqref{constant} in the opposite case $l_s/L \ll 1$. Different prescriptions remain possible, but go beyond the scope of this paper.

For $d\leq 3$, we obtain $l_s/L \ll 1$, for which we can use the $h_d$ values \eqref{constant}: for $d=3$ and $d=2$, we got $h_3 \simeq \tfrac{1}{2\pi}$ and $h_2 \simeq \tfrac{2}{\pi} \ln (L/l_s)$; This only gives little room to deviate from the standard spectrum \eqref{conddevstandardKK} in both cases, when combined with $l_s/L \ll 1$. Finally for $d=1$, the prescription \eqref{constantpresc1d} gives $h_1=L/l_s$ in an example of interest \eqref{constant1d}. Then, one will always get a deviation from the standard spectrum \eqref{conddevstandardKK}, and this will be physically relevant as long as $l_s/L \ll 1$. We now verify these claims for $d=1,2,3$, and get the precise deviations from the standard Kaluza--Klein spectrum \eqref{KKstandard}, thanks to a numerical evaluation.

\subsection{Numerical evaluation of the spectrum for $d=1,2,3$}\label{sec:num}

We evaluate numerically the Kaluza--Klein spectrum of gravitational waves propagating on a 4d warped Minkowski background with toroidal extra dimensions, by solving equation \eqref{intt7}. We use string motivated $D$-dimensional backgrounds, where the warp factor is due to objects analogue to $D$-branes and orientifolds, whose transverse space is a $d$-dimensional torus. The stringy input is the number and placement of orientifold-like sources (see e.g.~the background in section \ref{sec:p=3}), and the inspiration for the $D$-dimensional expression of their charges \eqref{QD}; the discussion of a physically relevant regime in section \ref{sec:KKscale} also uses such an input, e.g.~through the requirement $g_s<1$. One may still consider different models where the number of sources or their charges are different; for this reason our spectrum evaluation will go beyond a stringy physically relevant regime. We evaluate the spectrum only for $d=1,2,3$, for reasons displayed at the end of section \ref{sec:KKscale} but also because a higher $d$ would require increased computational power.

The tools and strategy relevant to this spectrum evaluation were introduced in section \ref{sec:KKscale}, with the operator $\ocal$ in \eqref{operator} and the text below. To proceed further it is convenient to consider the following operator
\beq
\ocal_{\vec{m}}{}^{\vec{p}} \times \frac{1}{g_s} \left(\frac{L}{l_s}\right)^{d-2}  =\ \ocal'_{\vec{m}}{}^{\vec{p}} = \left(\frac{\vec{m}^2}{\mu_N^2} -   \eta_d \right)\delta_{\vec{m}}^{\vec{p}} -
 \sum_{\vec{n}\in \mathbb{Z}^{d\, *}} \frac{\sum_i q_i e^{-2\pi \i \vec{n}\cdot \vec{\sigma_i}}}{ (2\pi\vec{n})^2}\delta_{\vec{m}-\vec{n}}^{\vec{p}} \ , \label{operatorprime}
\eeq
where we have defined
\eq{
 \eta_d = h_d\left(\frac{L}{l_s}\right)^{d-2} \ ,\quad \mu_N^2 = M_N^2 L^2  \times g_s \left(\frac{l_s}{L}\right)^{d-2}
~.}
The eigenvalue equation \eqref{intt7} can be rewritten in terms of the new parameters as
\eq{\label{inttt8}
-\delta^{mn} \frac{\partial}{\partial\sigma^m}\frac{\partial}{\partial\sigma^n} \psi_N= H'\, (2\pi\, \mu_N)^2\, \psi_N
~,}
with $H'$  given by
\eq{\label{hfnc}
H  \times \frac{1}{g_s} \left(\frac{L}{l_s}\right)^{d-2} =\ H' =- (2\pi L)^{d-2}\sum_i q_i\, G(\vec{y} - \vec{y}_i) + \eta_d
~.}
Provided the values $\mu_N$ are calculated by solving the system  $\ocal'_{\vec{m}}{}^{\vec{p}} c_{\vec{p}}=0$, the corresponding Kaluza--Klein spectrum $M_N$ is obtained, normalised to the standard Kaluza--Klein spectrum $\mathring{M}_N$ \eqref{KKstandard} in absence of sources; in other words we obtain the following ratio
\eq{
\frac{M_N}{\mathring{M}_N}=f_N~,\quad {\rm with}\ f_N=\frac{\sqrt{\eta_d}}{N}\, \mu_N  \label{fN}
~.}
The quantities $f_N$ are phenomenologically interesting as they measure the deviation from the standard spectrum \eqref{KKstandard}.

Without sources, the eigenfunctions for $d=1$ can be decomposed into odd and even functions, namely $\psi_N \propto \sin(2\pi N \sigma)$ and $\psi_N \propto \cos(2\pi N \sigma)$; in higher $d$, this gets generalized to antisymmetric ($a$) or symmetric ($s$) functions under $\vec{\sigma} \rightarrow - \vec{\sigma}$. There is thus a degeneracy in the standard Kaluza--Klein spectrum, that we denote as $\mu^{a}_N=\mu^{s}_N$. We will observe that this degeneracy gets lifted after the inclusion of sources, i.e.~with a non-constant warp factor, with typically $\mu^{a}_N < \mu^{s}_N$, and we will distinguish both spectra. In a stringy context, this becomes even more important because of the orientifold projection: the metric being even under an orientifold involution, the odd or antisymmetric eigenfunctions should eventually be projected-out. We will come back to this point in the following.

In the following we list numerical results concerning the first few Kaluza--Klein modes in dimensions $d=1,2,3$. On top of the number and location of the sources, the operator $\ocal'$ \eqref{operatorprime}, and therefore the spectrum, depends a priori on $\eta_d= h_d\left(L/l_s\right)^{d-2}$: we take for $h_d$ the values obtained in section \ref{sec:normH} and specify the value of the ratio $L/l_s$. We then present the antisymmetric and symmetric spectrum.

\subsubsection*{Case $d=1$}

We consider 2 $O_8$ placed at $\sigma=0, \tfrac{1}{2}$ and 16 $D_8$ all placed at $\sigma= 0$, such that the total charge vanishes: $\sum_i Q_i=0$. This configuration has been discussed around \eqref{constant1d}: from the prescription there, we deduce that $\eta_{1}=1$ is independent of the $l_s/L$ ratio, making it a special case. The spectrum is then unique and given in Table \ref{tab:specd=1}. It is obtained by solving the linear system corresponding to the operator $\ocal'$ \eqref{operatorprime}, truncating $c_{\vec{m}}=0$ for $|\vec{m}|> m_0=20$, following the strategy explained below \eqref{operator}. However the system of equations converges remarkably fast: truncating at $m_0=1$, the error compared to the truncation at $m_0=20$ is of order $10\%$; truncating at $m_0=2$, the same error is only of order $0.01\%$. Last but not least, we present in appendix \ref{ap:num} an alternative numerical method for $d=1$ to determine the eigenvalues, that goes back to the work of Hartree \cite{Hartree}: the spectrum of Table \ref{tab:specd=1} is reproduced.
\begin{table}[h]
  \begin{center}
    \begin{tabular}{|c||c|c||c|c|}
    \hline
   $N$ & \multicolumn{2}{c||}{$1$} & \multicolumn{2}{c|}{$2$} \\
   \hhline{-||--||--}
  $a/s$  & $a$ & $s$ & $a$ & $s$ \\
   \hhline{=====}
   $\mu_N$& 0.9800 & 1.1397 & 2.0368 & 2.2047 \\
     \hhline{-||--||--}
     $f_N$& 0.9800 & 1.1397 & 1.0184 & 1.1024 \\
      \hline
    \end{tabular}
     \caption{The first modes of the Kaluza--Klein spectrum for $d=1$, in terms of their deviation $f_N$ \eqref{fN} from the standard spectrum.}\label{tab:specd=1}
  \end{center}
\end{table}

\begin{figure}[H]
\begin{center}
\begin{subfigure}[H]{0.4\textwidth}
\includegraphics[width=\textwidth]{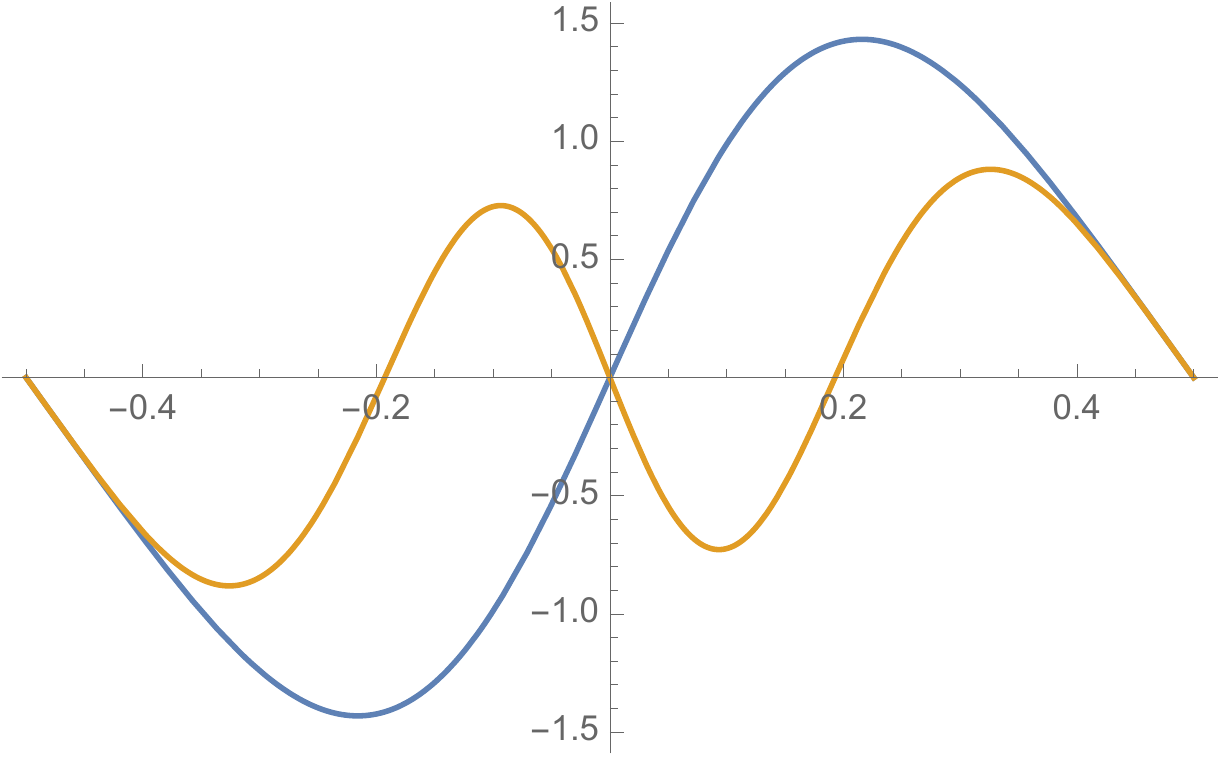}\caption{Odd eigenfunctions corresponding to $\mu^a_1$ (2 zeros) and $\mu^a_2$ (4 zeros)}\label{fig:o1}
\end{subfigure}
\qquad \qquad
\begin{subfigure}[H]{0.4\textwidth}
\includegraphics[width=\textwidth]{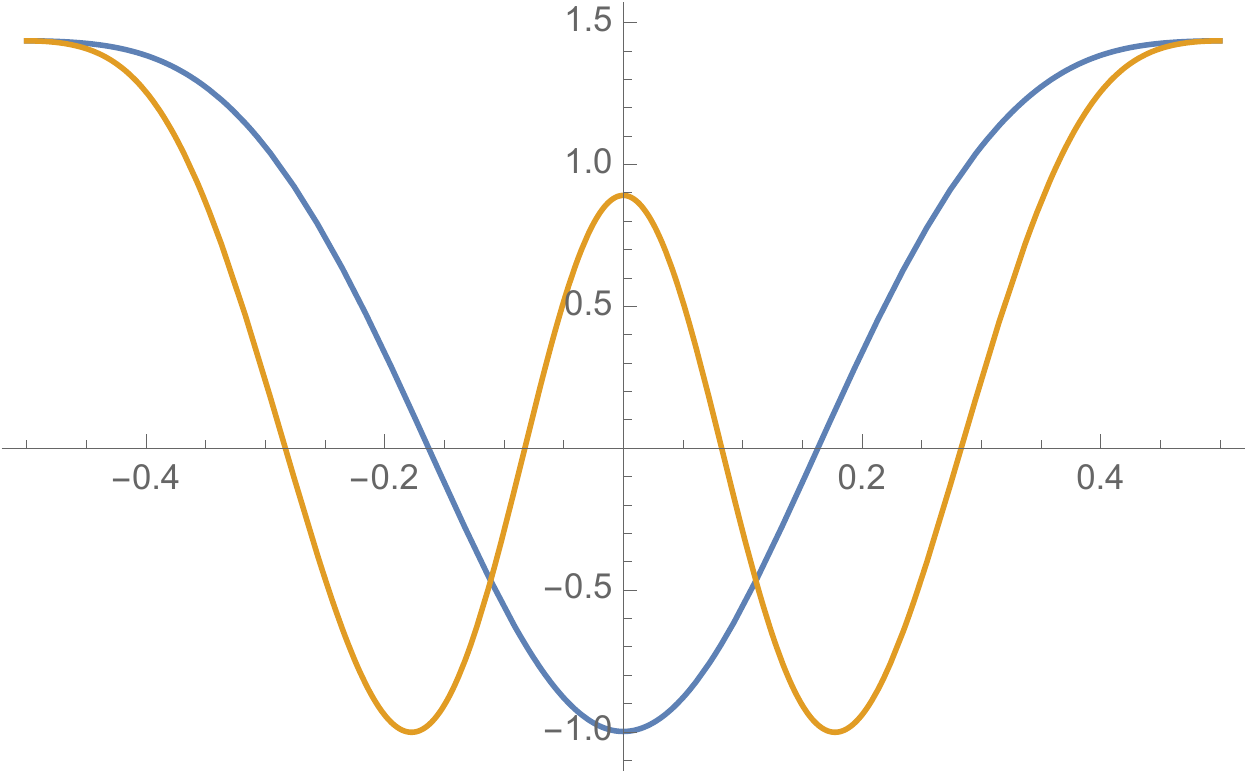}\caption{Even eigenfunctions corresponding to $\mu^s_1$ (2 zeros) and $\mu^s_2$ (4 zeros)}\label{fig:e1}
\end{subfigure}
\caption{Odd and even eigenfunctions corresponding to the eigenvalues of Table \ref{tab:specd=1}.}\label{fig:f1}
\end{center}
\end{figure}

In addition to the Kaluza--Klein eigenvalues, the solution of the system \eqref{operatorprime} allows us to determine the Fourier coefficients of the corresponding eigenfunctions. Figure \ref{fig:f1} depicts the eigenfunctions corresponding to the eigenvalues of Table \ref{tab:specd=1}. As expected from general theory, an $N$-th eigenfunction $\psi_N$ has $2N$ zeros.
\noindent We recall from the above general discussion that the orientifold would project out the odd eigenfunctions, leaving only the $s$ entries in Table \ref{tab:specd=1}.

As can be seen from the $f_N$ in Table \ref{tab:specd=1}, the spectrum is different than the standard Kaluza--Klein spectrum \eqref{KKstandard}, as anticipated at the end of section \ref{sec:KKscale}. But the deviation remains mild: this can be understood from \eqref{conddevstandardKK}, where the value $h_1=L/l_s$ only saturates the inequality.

\subsubsection*{Case $d=2$}

We consider 4 $O_7$ placed at $\vec{\sigma}=(0,0), (0,\tfrac{1}{2}), (\tfrac{1}{2},0), (\tfrac{1}{2},\tfrac{1}{2})$ and 16 $D_7$ all placed  at $\vec{\sigma}= \vec{0}$, so that the total charge vanishes. Here we have $\eta_{2}=h_2=\frac{2}{\pi}\ln(\tfrac{L}{l_s})$. The operator $\ocal'$ and thus the spectrum now depend on $L/l_s$, and we consider below different values for this ratio.\footnote{\label{foot:constant}We took for $h_2$ the value \eqref{constant} computed from the prescription \eqref{constantpresc} in the approximation $L/l_s \gg 1$. Some values of this ratio considered here are very mildly in that regime, in which case the same value for $h_2$ is still used but may not correspond to the previous prescription anymore.} The spectrum is obtained by truncating $c_{\vec{m}}=0$ at $m_0=2\sqrt{2}$. Its is given in Table \ref{tab:specd=2}. We see that the truncation only captures the first two antisymmetric eigenfunctions. In contrast to the $d=1$ case, the truncation we have imposed, for computational reasons, might be too restrictive, and causing the program to miss some of the low-lying symmetric modes, namely the first symmetric eigenfunctions. Another possibility would be that the latter correspond to higher eigenvalues. In any case, their absence is not very satisfactory if one imposes further the orientifold involution that projects out the antisymmetric eigenfunctions.
\begin{table}[h]
  \begin{center}
    \begin{tabular}{|c||c||c|c||c|c|}
    \hline
  & $N$ & \multicolumn{2}{c||}{$1$} & \multicolumn{2}{c|}{$\sqrt{2}$} \\
    \hhline{~-||--||--}
 $L/l_s$ & a/s & $a$ & $s$ & $a$ & $s$ \\
    \hhline{======}
  & $\mu_N$ & 1.3946 &  & 3.9954 &  \\
    \hhline{~-||--||--}
 1 & $f_N$   & 0 & \phantom{1.5477} &  0 & \phantom{1.5477} \\
    \hhline{======}
  & $\mu_N$ & 1.2196 &  & 2.8020 &  \\
    \hhline{~-||--||--}
 1.5 & $f_N$   & 0.6196 & \phantom{1.5477} & 1.0066 & \phantom{1.5477} \\
    \hhline{======}
  & $\mu_N$ & 1.1196 &  & 2.3298 &  \\
    \hhline{~-||--||--}
 2 & $f_N$   & 0.7438 & \phantom{1.5477} & 1.0944 & \phantom{1.5477} \\
    \hhline{======}
   & $\mu_N$ & 0.7804 &  & 1.2211 &  \\
    \hhline{~-||--||--}
 10 & $f_N$   & 0.9448 &  & 1.0454 &  \\
    \hhline{======}
  & $\mu_N$ & 0.5749 &  & 0.8357 &  \\
    \hhline{~-||--||--}
 $10^2$ & $f_N$   & 0.9844 &  & 1.0118 &  \\
    \hhline{======}
  & $\mu_N$ & 0.4735 &  & 0.6778 &  \\
    \hhline{~-||--||--}
 $10^3$ & $f_N$   & 0.9930 &  & 1.0051 &  \\
    \hhline{======}
  & $\mu_N$ & 0.4113 &  & 0.5857 &  \\
    \hhline{~-||--||--}
 $10^4$ & $f_N$ & 0.9960 &  & 1.0028 &  \\
     \hline
    \end{tabular}
     \caption{The first modes of the Kaluza--Klein spectrum for $d=2$, in terms of their deviation $f_N$ \eqref{fN} from the standard spectrum. An empty cell means that the mode has not been found within the search range.}\label{tab:specd=2}
  \end{center}
\end{table}

With $h_2=\frac{2}{\pi}\ln(\tfrac{L}{l_s})$ and the conditions \eqref{condstandardKK} or \eqref{conddevstandardKK}, we deduce that the bigger $L/l_s$ the closer the spectrum is to the standard Kaluza--Klein spectrum (while remaining in a physically relevant regime \eqref{condphysrel}). We verify this with the $a$ modes in Table \ref{tab:specd=2}, as the $f^a_N$ get closer to $1$ as $L/l_s$ increases. On the contrary, $L/l_s = 2$ giving $h_2 \simeq 0.44$ is an example of the situation mentioned at the end of section \ref{sec:KKscale}, where one gets a mild deviation from the standard spectrum, verified here with $f_1^a$, while being still barely in a physically relevant regime \eqref{condphysrel}. We cannot go lower in the ratio values than $L/l_s=1$, due to the expression used here for $h_2$. In that limit, we still obtain finite $\mu_N$, and we reach the maximal deviation from the standard spectrum with $f_N=0$.

\subsubsection*{Case $d=3$}

\begin{figure}[H]
\begin{center}
\includegraphics[width=0.7\textwidth]{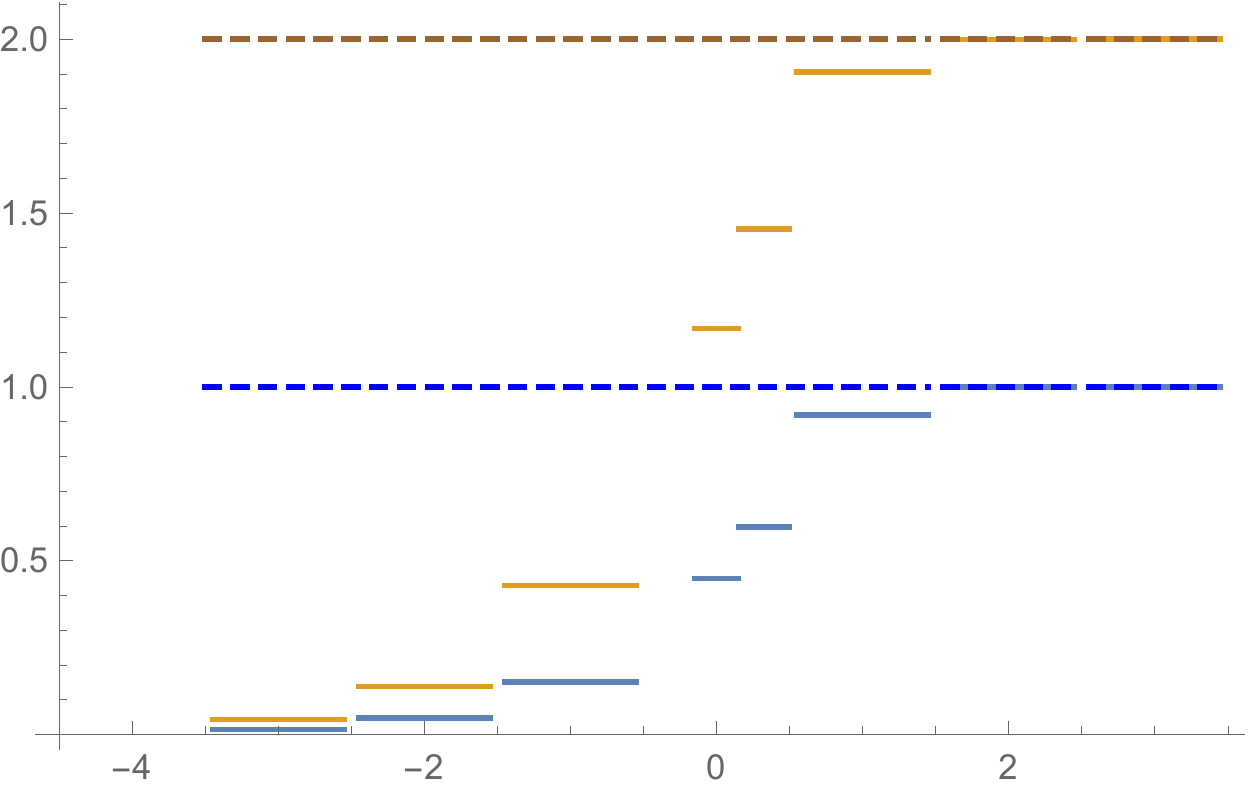}
\caption{Kaluza--Klein spectrum $M_N^2 \, (\times H_0 L^2)$ with $d=3$ for a non-trivial warp factor (plain lines), compared to the standard spectrum for a constant warp factor (dashed lines), depending on the parameter value $\log_{10}(L/{l_s})$. The bottom (blue) lines correspond to $a$-modes with $N=1$, the upper lines to $s$-modes with $N=\sqrt{2}$, as given in Table \ref{tab:specd=3}.}\label{fig:spectre}
\end{center}
\end{figure}

We consider 8 $O_6$ placed at $\vec{\sigma}=(0,0,0), (0,0,\frac{1}{2}),  \dots, (\frac12,\frac{1}{2},\frac12)$ and 16 $D_7$ all placed  at $\vec{\sigma}= \vec{0}$, so that the total charge vanishes. We have $\eta_{3}=\frac{1}{2\pi}\tfrac{L}{l_s}$. As for $d=2$, the operator $\ocal'$ and the spectrum depend on $L/l_s$, for which we then consider different values.\footnote{The same remark as in footnote \ref{foot:constant} can be made on the $h_3$ value and the prescription \eqref{constantpresc} to compute it.} The spectrum is obtained by truncating $c_{\vec{m}}=0$ at $m_0=\sqrt{3}$. Its is given in Table \ref{tab:specd=3}, and illustrated in Figure \ref{fig:spectre}. Contrary to $d=2$, the truncation, if to be trusted, captures some symmetric eigenfunctions, even though not the first ones. Those would remain after imposing the orientifold projection.

For $d=2,3$, we had to identify the $N$ for each mode. This $N$ is the level of a corresponding mode in the standard spectrum, to which we compare. For high $L/l_s$ ratios, we are sufficiently close to the standard spectrum to allow for such an identification, as indicated in \eqref{condstandardKK}: we can either find $N$ such that $f_N \simeq 1$, or evaluate Fourier coefficients of an eigenfunction and compare it to products of $cos$ and $sin$, or also count the number of zeros of the eigenfunction along each direction. Getting to lower $L/l_s$ makes any comparison to standard spectrum modes, and the $N$ identification, more difficult. It is thus the continuity of the $\mu_N$ and $f_N$ values starting from high $L/l_s$ that guides us in the identification of the modes. The order in which the eigenmodes appear may also help, even though as explained, the truncation may lead to miss some of them.
\begin{table}[h]
  \begin{center}
    \begin{tabular}{|c||c||c|c||c|c|}
    \hline
  & $N$ & \multicolumn{2}{c||}{$1$} & \multicolumn{2}{c|}{$\sqrt{2}$} \\
   \hhline{~-||--||--}
 $L/l_s$ & a/s & $a$ & $s$ & $a$ & $s$ \\
    \hhline{======}
  & $\mu_N$ & 1.2075  &   \phantom{1.2075}  && 2.4515  \\
    \hhline{~-||--||--}
 $0$ & $f_N$ & 0 & & &0    \\
    \hhline{======}
     & $\mu_N$ & 1.2074 & &&2.4510   \\
    \hhline{~-||--||--}
 $10^{-3}$ & $f_N$ & 0.0152 & && 0.0218  \\
    \hhline{======}
  & $\mu_N$ & 1.2066 & && 2.4466   \\
    \hhline{~-||--||--}
 $10^{-2}$ & $f_N$ &  0.0481& && 0.0690 \\
    \hhline{======}
  & $\mu_N$ & 1.1988 & &&2.4040   \\
    \hhline{~-||--||--}
 $10^{-1}$ & $f_N$ & 0.1512 &&& 0.2145   \\
    \hhline{======}
       & $\mu_N$ & 1.1265 & &&2.0706   \\
    \hhline{~-||--||--}
 $1 $ & $f_N$ & 0.4494 & &&0.5841  \\
    \hhline{======}
  & $\mu_N$ & 1.0576 &&& 1.8223   \\
    \hhline{~-||--||--}
 2 & $f_N$   & 0.5967 & &&0.7270  \\
    \hhline{======}
   & $\mu_N$ & 0.7288 & & 1.1168& 1.0685   \\
    \hhline{~-||--||--}
 10 & $f_N$   & 0.9194 &&0.9962& 0.9531  \\
    \hhline{======}
  & $\mu_N$ & 0.2504 & & 0.3515 &0.3543   \\
    \hhline{~-||--||--}
 $10^2$ & $f_N$   & 0.9991 & &0.9914&0.9993    \\
   \hhline{======}
  & $\mu_N$ & 0.0793 & & 0.1120&0.1121   \\
    \hhline{~-||--||--}
 $10^3$ & $f_N$   & 1.0000 & &0.9991&1.0000   \\
     \hline
    \end{tabular}
     \caption{The first modes of the Kaluza--Klein spectrum for $d=3$, in terms of their deviation $f_N$ \eqref{fN} from the standard spectrum. An empty cell means that the mode has not been found within the search range.}\label{tab:specd=3}
  \end{center}
\end{table}

With both the symmetric and antisymmetric modes, we verify with the $f_N$ in Table \ref{tab:specd=3} or the illustration in Figure \ref{fig:spectre} the behaviour discussed at the end of section \ref{sec:KKscale}: the bigger $L/l_s$, the closer we get to the standard Kaluza--Klein spectrum, and the safer we are in terms of a physically relevant regime, at least in a stringy framework. $L/l_s=2$ is again close to the boundary value on these aspects. On the contrary, the smaller $L/l_s$, the bigger the deviation from the standard spectrum. This holds until the limit of maximal deviation with $f_N=0$, where we verified that the $\mu_N$ reach a finite value. Justifying the physical relevance of the parameter values with small $L/l_s$ would amount to consider a different framework or model, e.g.~with different source charges, as discussed at the beginning of this section. Let us emphasize that the smaller $L/l_s$ values lead to the phenomenologically most interesting deviations from the standard Kaluza--Klein spectrum: the $f_N$ get indeed much smaller than 1, i.e.~the first Kaluza--Klein masses get noticeably lowered, bringing them closer to observability bounds.

\section{Summary and discussion}\label{sec:ccl}

In this work, we have studied 4d gravitational waves propagating on a $D$-dimensional background made of a 4d warped Minkowski space-time and $D-4$ extra dimensions. The latter lead to a tower of Kaluza--Klein 4d gravitational waves, whose mass spectrum is affected by the non-trivial warp factor $H$, as can be seen in the eigenvalue equation \eqref{eigeneq}. Contrary to braneworld models mentioned in the Introduction, the warp factor is due to $D_p$-branes and orientifold $O_p$-planes sources, or analogous $p$-dimensional objects in $D$-dimensions, and determined as a solution to a Poisson equation \eqref{BI}. As such, $H$ is given in terms of generalized Green's functions $G$, the source charges $Q_i$, and a constant $H_0$, as in \eqref{H}. These various ingredients are however typically not given explicitly for compact extra dimensions, so we determine them here in section \ref{sec:Greengal} for a transverse $d$-dimensional torus $\mathbb{T}^d$, $d=D-p-1$. We provide an expression for $G$ in \eqref{a3}, inspired by Courant-Hilbert \cite{Courant:1989aa} and proposals of\cite{Shandera:2003gx, Kim:2018vgz}; illustrations of $G$ are provided in Figure \ref{fig:G}. This expression passes further non-trivial tests: in appendix \ref{ap:green}, we reproduce analytically from it the expected behaviour close to the source, and indicate a matching to a different expression known for $d=2$ in the context of string amplitudes. We turn to the constant $H_0$ which plays a crucial role: it corresponds to the average of $H$ over $\mathbb{T}^d$, in the case $\sum_i Q_i=0$ mostly considered here.
It is also required to be non-zero to guarantee a standard massless gravitational wave, and to maintain in addition $H\geq 0$, avoiding a signature change in the extra dimensions. Based on the latter, we propose a prescription \eqref{constantpresc} that fixes $H_0 = g_s h_d$, and compute $h_d$ in some cases. Resulting warp factors are depicted in Figure \ref{fig:H3d} and \ref{fig:H1d}. We turn in section \ref{sec:spectrumgal} to the analysis of the 4d gravitational waves Kaluza--Klein spectrum on such warped backgrounds: those include string compactifications, but our formulation in $D$-dimensions with free parameters allows for more general models. The standard Kaluza--Klein spectrum for a trivial, i.e. constant, warp factor is given in \eqref{KKstandard} (see also footnote \ref{foot:intro}). We analyse in section \ref{sec:KKscale} what values of the parameters $L/l_s$ and $h_d$ allow for a deviation from this spectrum \eqref{conddevstandardKK}, while remaining in a (stringy motivated) physically relevant regime \eqref{condphysrel}; $L$ is the radius of the unwarped $\mathbb{T}^d$ and $l_s$ is the fundamental length, e.g.~string length. We finally evaluate numerically this spectrum for $d=1,2,3$, with various values of $L/l_s$, and quantify the deviation from the standard spectrum by the number $f_N$ \eqref{fN} for each mode $N$. The degeneracy of the spectrum into symmetric ($s$) and antisymmetric ($a$) modes gets lifted. The spectrum and deviations obtained for the first few modes are given in Table \ref{tab:specd=1}, \ref{tab:specd=2} and \ref{tab:specd=3}, and the spectrum for $d=3$ is depicted in Figure \ref{fig:spectre}.

\subsubsection*{Observability of Kaluza--Klein gravitational waves}

As recalled in the Introduction, standard Kaluza--Klein spectra lead to gravitational waves of frequencies far too high to be detected by current ground-based observatories. However, such a detection  could be made possible in the future   with eLISA,  through the observation of the stochastic gravitational waves background (SGWB) of cosmological origin. Indeed, gravitational waves produced at high energies (in particular with high mass or frequency) in the early universe would benefit from the cosmological redshift to be observed today \cite{Caprini:2018mtu, Caprini:2019pxz}. It would be interesting to have more precise bounds on the frequency and amplitude for such detectable primordial Kaluza--Klein gravitational waves.

Coming back to our results, the first point is that we observe a deviation due to a non-trivial warp factor with respect to the standard spectrum: this deviation is even independent of the parameter $L/l_s$ for $d=1$, and present for some range of values of $L/l_s$ for $d=2,3$. Therefore, if Kaluza--Klein gravitational waves are ever observed, the discrepancy of these spectra is interesting. Secondly, in physically relevant regimes for a string framework \eqref{condphysrel}, namely for $L/l_s \gtrsim 10$, the deviation is unfortunately not big for the cases considered: we obtained that the lowest Kaluza--Klein mass is at best a half of the standard one. The fact the mass gets lowered with a non-trivial warp factor is encouraging, but so far not enough to be observed. There are however few caveats or possible improvements in this answer to our initial question, regarding the constant $H_0$, the dimensions $d$ considered and the need of more computational power; we come back to those at the end of this section. In addition, let us emphasize that different parameter regimes, e.g.~$L/l_s \ll 1$ for $d=3$, lead to huge deviations and lowering of the masses. Such regimes could maybe be justified in different models, or with different prescriptions for the constant $H_0$; further possibilities are again mentioned below.

\subsubsection*{An explicit warp factor for string compactifications}

Another important result in this paper is to provide an explicit expression for a warp factor, that applies (in particular) to string compactifications. As mentioned in the Introduction, this is usually not done explicitly, especially to the extent of solving generalized Green's functions and fixing the constant $H_0$. Even though one may argue that phenomenological cases of interest go beyond a torus $\mathbb{T}^d$ or $\sum_i Q_i=0$ as considered here, several important ideas and constructions in string phenomenology could already be tested thanks to this explicit expression. We list here a few, hoping to come back to them in the future:\\
- The physics close to the sources: it plays a crucial role in recent discussions on de Sitter solutions, e.g.~in \cite{Cordova:2018dbb, Cribiori:2019clo} that we relate here to Figure \ref{fig:H1d}, or through the study of local physics in the throats \cite{Bena:2018fqc, Carta:2019rhx, Blumenhagen:2019qcg, Bena:2019sxm} (see also \cite{Frey:2006wv, Burgess:2006mn}). In particular, the discussion related to the prescription on $H_0$ \eqref{constantpresc} could be relevant: there we propose a horizon cut close to orientifolds to maintain $H>0$. A similar cut should be imposed close to $D$-branes to maintain a finite integral of $H$, i.e.~prevent from an infinite volume and infinite internal distances. This is reminiscent of singularity resolutions in throats.\\
- The validity of the smearing approximation: see e.g.~\cite{Blaback:2010sj, Junghans:2013xza} and some possible implications for de Sitter solutions in \cite{Das:2019vnx}. Here for instance, the warp factor would replaced by its average (constant) value and sources contributions by the $C$ constant.\\
- Effective field theories with warp factors: formal results can be found in \cite{Giddings:2005ff, Shiu:2008ry, Douglas:2008jx, Frey:2008xw, Martucci:2009sf, Martucci:2014ska, Grimm:2014efa, Grimm:2015mua}, in which one could now replace the warp factor by its explicit expression.

\subsubsection*{Going further}

To get a lower Kaluza--Klein spectrum, an idea is to introduce a different scale in the problem, bigger in length than the average internal radius. This is a standard question for moduli stabilization in string theory, or in attempts to obtain interesting phenomenology, and usual possibilities involve fluxes and curvature. Here, these options would translate into having a $C= \tfrac{1}{V}\sum_i Q_i \neq 0$, or the first non-zero eigenvalues of the Laplacian small compared to the inverse internal radius. This last situation occurs for one-forms on nilmanifolds as shown in \cite{Andriot:2018tmb}: the smallest non-zero eigenvalue is related to the curvature of the manifold, which introduces a different scale than the average radius. It would then be interesting to analyse the warp factor and spectrum on a curved manifold instead of the torus, or with $C\neq 0$. Let us also recall that the average of $H$ is modified if $C \neq 0$, which makes the interpretation of $H_0$ more difficult. So far though, few numerical tests have shown little difference on the spectrum for $C \neq 0$, but we hope to examine these ideas more thoroughly in the future.

Finally, it would be interesting to combine effects of the warp factor to that of the other fields coming from extra dimensions, namely the vectors and scalars as discussed in the Introduction. The setting considered here, 4d TT (Kaluza--Klein) gravitational waves, is known to be a consistent truncation of the fully fluctuated $D$-dimensional theory \cite{Andriot:2017oaz}, as recalled in section \ref{sec:1mg}. This should help to extend the study to the other fields. In particular, it would be interesting to see how the effect on the polarization due to the massless scalar field \cite{Andriot:2017oaz} is modified. One may also wonder whether the Kaluza--Klein spectrum of the remaining fields is altered.

\subsubsection*{More on the spectrum evaluation}

Last but not least, we end here with some technical points that could improve the current spectrum evaluation, or help to get new spectra, e.g.~with $d=4,5,6$ or a different constant $H_0$. The spectrum has been evaluated by a numerical resolution of the eigenvalue equation \eqref{intt7} or \eqref{inttt8}, following a method presented in section \ref{sec:KKscale} and \ref{sec:num} (see also appendix \ref{ap:num} for $d=1$). It required a truncation to an integer $m_0$ (for a finite size matrix) that appeared very reliable for $d=1$, but maybe more problematic for $d=2,3$. For the latter, it might cause in particular to miss some of the low-lying Kaluza--Klein modes, as discussed in section \ref{sec:num}. A first improvement would be to either find a different resolution method, or more computational power to allow for a bigger $m_0$ and a better control on the truncation.

More computational power would also be needed to evaluate the spectrum for $d=4,5,6$. The cases $d=5,6$ especially could lead to very interesting results as argued at the end of section \ref{sec:KKscale}. On general grounds, we expect the deviation from the standard Kaluza--Klein spectrum to be much bigger for $d=5,6$ because the non-constant part of the warp factor is much stronger: see e.g.~Figure \ref{fig:G}.

Finally, as pointed out at the end of section \ref{sec:KKscale}, the value of the constant $H_0$ is crucial. The prescription \eqref{constantpresc} we proposed is a minimal choice of $H_0$ to guarantee $H \geq 0$: $H$ vanishes at a horizon distance from the source (or at the source for $d=1$). One could consider a different prescription, in particular one making $H$ more positive, i.e.~a bigger $H_0$.\footnote{This apparent freedom in $H_0$ can conceptually be related to the fact we have an unfixed volume modulus, i.e.~the size of the $d$ transverse directions is not fixed in our setting, even though the latter also involves $L$.} Having a big $H_0$ or $h_d$ however does not seem to help in terms of deviations from the standard Kaluza--Klein spectrum, as can be seen in \eqref{conddevstandardKK}, but a bigger value than the one computed in \eqref{constant} appears nevertheless necessary for $d=5,6$ as mentioned at the end of section \ref{sec:KKscale}. Before going to a different prescription for $H_0$, we should also recall that we computed $h_d$ only for $L/l_s \gg 1$ in \eqref{constant}, and we used these values all along; it would be interesting to first compute $h_d$ from our prescription \eqref{constantpresc} in different regimes. These various technical points are simple ideas to improve and extend the spectrum evaluation and we hope to come back to them in the future.

\vspace{0.4in}

\subsection*{Acknowledgements}

We warmly thank A.~Kiefer, S.~Sethi and P.~Tourkine for helpful exchanges during the completion of this work. D.~A.~acknowledges support from the Austrian Science Fund (FWF): project number M2247-N27.

\newpage

\begin{appendix}

\section{Study of the generalized Green's function}\label{ap:green}

We proposed in section \ref{sec:Green} the expression \eqref{a3} for the generalized Green's function $G$ on a $d$-dimensional torus, following a prescription by \cite{Shandera:2003gx, Kim:2018vgz}. This expression allows us to reproduce a rigorous result by Courant-Hilbert \cite{Courant:1989aa} for $d=3$, presented in section \ref{sec:threebrane}. We study here this expression further by first determining the behaviour of $G$ close to the source, and then focusing on the $d=2$ case for which another expression is known.

\subsection{Behaviour close to the source}\label{ap:behaviour}

We expect the behaviour of $G(\vec{\sigma})$ \eqref{a3} close to the source at $\vec{\sigma}=\vec{0}$ to match the one obtained in a non-compact flat space. To verify this, we first introduce a primitive function (indefinite integral) $F(\vec{\sigma},t)$ of $1-\prod_{m=1}^d\theta_3(\sigma^m|4\pi \i t)$ with respect to $t$ such that formally, i.e.~if everything is well-defined, one gets
\beq
(2\pi L)^{d-2}\ G(\vec{\sigma})  = F(\vec{\sigma},\infty) - F(\vec{\sigma},0) \ .
\eeq
We first study the behaviour at $t \rightarrow \infty$: one has
\beq
\left|1-\prod_{m=1}^d\theta_3(\sigma^m|4\pi \i t)\right| = \left|\sum_{\vec{n}\in \mathbb{Z}^{d\, *}} e^{2\pi \i \vec{n}\cdot \vec{\sigma}-4\pi^2\vec{n}^2 t } \right| \leq \left(\sum_{n\in \mathbb{Z}^*}  e^{-4\pi^2 n^2 t } \right)^d \leq \left(\sum_{n\in \mathbb{Z}^*}  e^{-4\pi^2 n t } \right)^d
\eeq
i.e.
\beq
\left|1-\prod_{m=1}^d\theta_3(\sigma^m|4\pi \i t)\right| \leq \left(\frac{ e^{-4\pi^2 t }}{1- e^{-4\pi^2 t } } \right)^d  \ \sim_{t\rightarrow \infty}\ e^{-4\pi^2 d t } + o(e^{-4\pi^2 d t }) \ .
\eeq
We deduce that $-e^{-4\pi^2 d t } \leq 1-\prod_{m=1}^d\theta_3(\sigma^m|4\pi \i t) \leq e^{-4\pi^2 d t }$ up to terms $o(e^{-4\pi^2 d t })$, and since primitives on both sides go to zero at $t \rightarrow \infty$, independently of $\vec{\sigma}$, we conclude
\beq
F(\vec{\sigma},\infty) = 0 \ .
\eeq
There is no definition issue on this side of the integral. We turn to the other side, $t\rightarrow 0$. Let us first recall properties of the $\theta_3$-function \eqref{thetadef}
\eq{\label{a5}
\theta_3(-\sigma|\tau)=\theta_3(\sigma|\tau)~,~~~
\theta_3(\sigma+1|\tau)=\theta_3(\sigma|\tau)~,~~~\theta_3(\sigma|\tau)=\frac{e^{-\frac{\i \pi \sigma^2}{\tau}}}{(-\i \tau)^{\frac{1}{2}}} \theta_3\left(-\frac{\sigma}{\tau}\big|-\frac{1}{\tau}\right)
~.
}
Thanks to the last property, namely modularity, on $\tau= \i t$, one obtains
\beq
\theta_3(\sigma|\i t)=\frac{e^{-\frac{\pi \sigma^2}{t}}}{\sqrt{t}} \left( 1+ 2 \sum_{n=1}^\infty e^{-\frac{\pi n^2}{t}} \cosh \left(\frac{2\pi n \sigma}{t}\right)\right) = \frac{1}{\sqrt{t}} \sum_{n\in \mathbb{Z}} e^{- \frac{\pi}{t}(\sigma + n)^2} \ ,
\eeq
where we clearly see the periodicity in $\sigma$. Restricting for our purposes to $\sigma  \in [-\tfrac{1}{2},\tfrac{1}{2}]$, we deduce from the above the behaviour when $t \rightarrow 0$
\beq
\theta_3(\sigma|4\pi \i t)\ \sim_{t\rightarrow 0}\ \frac{ e^{- \frac{\sigma^2}{4 t}}}{\sqrt{4\pi t}} \ . \label{thetat0}
\eeq
This term dominates any other in the sum, except on the boundaries $\sigma = \pm \frac{1}{2}$ where it becomes equal to the neighbouring terms: this is expected from continuity and periodicity. The following should then be understood away from $\sigma = \pm \frac{1}{2}$, in particular close to $\sigma =0 $. We infer that
\beq
(2\pi L)^{d-2}\ G(\vec{\sigma}) = - F(\vec{\sigma},0) \sim_{\vec{\sigma}^2 \rightarrow 0}\  \frac{1}{4 \pi^{\frac{d}{2}}} \ {\rm Primitive}\!\left[ \frac{1}{T^{\frac{d}{2}}} e^{-\frac{\vec{\sigma}^2}{T}} \right]\!(T=0) \ . \label{exprGapp}
\eeq
Since $e^{-\frac{\vec{\sigma}^2}{T}} \sim_{T\rightarrow \infty} \, 1$ for any $\vec{\sigma}$, one has ${\rm lim}_{T\rightarrow \infty}\, {\rm Primitive}\!\left[ \frac{1}{T^{\frac{d}{2}}} e^{-\frac{\vec{\sigma}^2}{T}} \right] = {\rm lim}_{T\rightarrow \infty}\, T^{1-\frac{d}{2}} = 0$ for $d>2$, so we rewrite the above as
\beq
d>2:\quad  (2\pi L)^{d-2}\ G(\vec{\sigma}) \sim_{\vec{\sigma}^2 \rightarrow 0}\  - \frac{1}{4 \pi^{\frac{d}{2}}} \int_{0}^{\infty} \d T\, \frac{1}{T^{\frac{d}{2}}} e^{-\frac{\vec{\sigma}^2}{T}} = - \frac{1}{4 \pi^{\frac{d}{2}}} \int_{0}^{\infty} \d u\, u^{\frac{d}{2}-2} e^{-\vec{\sigma}^2 u} \ ,
\eeq
for which we use known integrals. We obtain
\begin{empheq}[innerbox=\fbox, left=\!\!\!\!\!\!\!\!\!\!\!\!\!\!\!\!\!\!]{align} \label{behav3d}
d\geq 3 &:\quad  (2\pi L)^{d-2}\ G(\vec{\sigma}) \sim_{\vec{\sigma}^2 \rightarrow 0}\  - \frac{1}{4 \pi^{\frac{d}{2}}}\ \Gamma\left(\tfrac{d-2}{2} \right)\, \frac{1}{|\vec{\sigma}|^{d-2}}
\end{empheq}
where for $d=3$ we used a different integral than for $d\geq 4$, but expressions eventually match.\footnote{A different method for $d=3$ was proposed in \cite{Blaback:2015gwa}.}

For $d=2$, we rewrite \eqref{exprGapp} as
\beq
d=2:\quad G(\vec{\sigma}) \sim_{\vec{\sigma}^2 \rightarrow 0}\  -\frac{1}{4 \pi} \ {\rm Primitive}\!\left[ \frac{1}{u} e^{-\vec{\sigma}^2 u} \right]\!(u \rightarrow \infty) \ .
\eeq
This value is however problematic in the limit $\vec{\sigma}^2 \rightarrow 0$; even bounds we can obtain to characterise the behaviour of the function are ill-defined in that limit, because contrary to the above, they are not independent of $\vec{\sigma}$. To capture the physics of this particular limit, an idea is to subtract possible divergences, by comparing to any finite constant value $u=\lambda > 0$, for any $\vec{\sigma}^2 \neq 0$ (possibly close to $0$). We get
\beq
d=2:\quad G(\vec{\sigma}) \sim_{\vec{\sigma}^2 \rightarrow 0}\  -\frac{1}{4 \pi} \ \int_{\lambda}^{\infty}\d u\, \frac{1}{u} e^{-\vec{\sigma}^2 u} = -\frac{1}{4 \pi} \ \int_{\lambda \vec{\sigma}^2}^{\infty}\d v\, \frac{1}{v} e^{- v} = -\frac{1}{4 \pi} E_1(\lambda \vec{\sigma}^2) \ ,
\eeq
where $-E_1(-x)=E_i(x)$ is the exponential integral function. Using its series expansion, one obtains for $\vec{\sigma}^2 \sim 0$, with $\gamma$ being the Euler-Mascheroni constant,
\beq
G(\vec{\sigma}) \sim_{\vec{\sigma}^2 \rightarrow 0}\ \frac{1}{4 \pi} (\gamma + \ln |\lambda \vec{\sigma}^2| - \lambda \vec{\sigma}^2 + o(\vec{\sigma}^2) ) \ .
\eeq
In other words, given the Green's function is defined up to a constant, we obtain the following dominant contribution
\begin{empheq}[innerbox=\fbox, left=\!\!\!\!\!\!\!\!\!\!\!\!\!\!\!\!\!\!]{align} \label{behav2d}
d=2 &:\quad G(\vec{\sigma}) \sim_{\vec{\sigma}^2 \rightarrow 0}\ \frac{1}{2 \pi} \ln |\vec{\sigma}|
\end{empheq}
For $d=1$, we proceed similarly: we first write
\beq
(2\pi L)^{-1}\ G(\vec{\sigma}) \sim_{\vec{\sigma}^2 \rightarrow 0}\  -\frac{1}{4 \pi^{\frac{1}{2}}} \ {\rm Primitive}\!\left[ \frac{1}{u^{\frac{3}{2}}} e^{-\vec{\sigma}^2 u} \right]\!(u \rightarrow \infty) =  \frac{\vec{\sigma}^2}{2 \pi^{\frac{1}{2}}} \ {\rm Primitive}\!\left[ \frac{1}{u^{\frac{1}{2}}} e^{-\vec{\sigma}^2 u} \right]\!(u \rightarrow \infty) \ ,\nn
\eeq
where we have done an integration by parts. We then introduce as above a finite value $u=\lambda$ and get
\bea
(2\pi L)^{-1}\ G(\vec{\sigma}) \sim_{\vec{\sigma}^2 \rightarrow 0} &\ \frac{\vec{\sigma}^2}{2 \pi^{\frac{1}{2}}} \ \int_{\lambda}^{\infty}\d u\, \frac{1}{u^{\frac{1}{2}}} e^{-\vec{\sigma}^2 u} = \frac{|\vec{\sigma}|}{\pi^{\frac{1}{2}}} \ \int_{\lambda |\vec{\sigma}|}^{\infty}\d v e^{-v^2} \\
& = \frac{|\vec{\sigma}|}{\pi^{\frac{1}{2}}} \left( \int_{0}^{\infty}\d v e^{-v^2} - \frac{\sqrt{\pi}}{2} {\rm erf}(\lambda |\vec{\sigma}|) \right) =  \frac{|\vec{\sigma}|}{2} \left( 1 -  {\rm erf}(\lambda |\vec{\sigma}|) \right) \ .\nn
\eea
Since the function ${\rm erf}$ is linear close to zero, we deduce the following dominant contribution
\beq
d=1 :\quad (2\pi L)^{-1}\ G(\vec{\sigma}) \sim_{\vec{\sigma}^2 \rightarrow 0}\ \frac{|\vec{\sigma}|}{2} \ . \label{behav1d0}
\eeq
While we complete below the study of the $d=1$ case, we conclude already that for all $d$, we recover the expected behaviour near the source. The cases $d=1,2$ have required the introduction of some cutoff to reach this conclusion.

We have studied generalized Green's functions up to constants, since those do not enter the defining differential equation \eqref{BIG}. For $d\geq 2$, additional constants in $G$ do not matter close to the source since the function is divergent there: the behaviour is then dominated by the divergence. However for $d=1$, the behaviour is finite: the expression given in \eqref{behav1d0} even vanishes at the source. An additional constant would then matter. For $d=1$, the Fourier series expression \eqref{GFourier} of the Green's function is actually well-defined since the sum converges; it does not suffer from the problem specified in \eqref{pbsum} for $d\geq 2$. This makes us confident that we can equally use the Fourier series expression for $d=1$. We compute this way the constant of interest: since \eqref{behav1d0} vanishes at the source, we compute it there and get $ (2\pi L)^{-1} \ G(0)= - 2 \sum_{n\in \mathbb{N}^{*}} \frac{1}{4\pi^2 n^2} = -\tfrac{1}{12} $. We compute similarly the following value: $ (2\pi L)^{-1} \ G(\pm \frac{1}{2})= - 2 \sum_{n\in \mathbb{N}^{*}} \frac{(-1)^n}{4\pi^2 n^2} = \tfrac{1}{24}$, for later convenience. We conclude
\begin{empheq}[innerbox=\fbox, left=\!\!\!\!\!\!\!\!\!\!\!\!\!\!\!\!\!\!]{align} \label{behav1d}
d=1 &:\quad (2\pi L)^{-1}\ G(\vec{\sigma}) \sim_{\vec{\sigma}^2 \rightarrow 0}\ -\frac{1}{12} + \frac{|\vec{\sigma}|}{2}
\end{empheq}

We illustrate this study by plotting in Figure \ref{fig:G} in section \ref{sec:Greengal} the Green's function \eqref{a3} for various dimensions: the different behaviours close to the source become apparent. We also see that the function is periodic at $\pm \tfrac{1}{2}$. Finally, we successfully check for $d=1$ the values at $0$ given by $-\tfrac{1}{12} \sim -0.083 $, and at $\pm \tfrac{1}{2}$ given by $\tfrac{1}{24}$. The slope of $\tfrac{1}{2}$ near the source is also well verified.

\subsection{The $d=2$ case}\label{ap:2d}

An expression for the generalized Green's function on a $d=2$ torus is known in the string theory literature. We follow \cite{Kiritsis:2007zza} with slightly different conventions than those used so far. We consider a torus with coordinates $\sigma^{i=1,2} \in [0,1]$, identified as $\sigma^i \sim \sigma^i +1$. The metric is $\d s^2 = g_{ij} \d \sigma^i \d \sigma^j$ with $\tau = \tau_1 + \i \tau_2$, $\tau_i$ being real, $\tau_2 >0$, and
\beq
g=\frac{1}{\tau_2} \left(\begin{array}{cc} 1 & \tau_1 \\ \tau_1 & |\tau|^2 \end{array}\right) \ ,\ g^{-1}=\frac{1}{\tau_2} \left(\begin{array}{cc} |\tau|^2 & -\tau_1 \\ -\tau_1 & 1 \end{array}\right) \ ,\ {\rm det}(g) = 1 \ .
\eeq
This matches our previous conventions for $2\pi L=1$ on our side and here $\tau=\i$. One can introduce the complex coordinate $z=\sigma^1 + \tau \sigma^2$ that admits the identifications $z\sim z+1 \sim z+\tau$. One rewrites $\d s^2= \tfrac{1}{\tau_2} |\d \sigma^1 + \tau \d \sigma^2 |^2$, and the Laplacian is $\Delta = g^{ij}\del_{\sigma^i} \del_{\sigma^j} = \tfrac{1}{\tau_2} |\tau \del_{\sigma^1} - \del_{\sigma^2}|^2$. The following equation is considered
\beq
\Delta G = \frac{1}{\tau_2} (\delta(\vec{\sigma}) - 1) \ ,
\eeq
which is at first solved using a Fourier series with $\vec{n} = (p,q)$, $\vec{n} \cdot \vec{\sigma} = p\sigma^1 + q \sigma^2$, and
\beq
G= - \sum_{\vec{n} \in \mathbb{Z}^{2\, *}} \frac{e^{2\pi \i \vec{n} \cdot \vec{\sigma}}}{4\pi^2 |\tau p - q|^2} \ , \label{Fourier2d}
\eeq
up to a constant. As explained in \eqref{pbsum}, the sum is not absolutely convergent, and one may consider a $\zeta$-function regularization. The final generalized Green's function is given by
\beq
\tilde{G}= \frac{1}{2\pi} \ln\left| \theta_1 (z|\tau) \right| - \frac{({\rm Im}(z))^2}{2 {\rm Im}(\tau)} + f(\tau) \ , \label{G2d}
\eeq
where coefficients are usually adjusted according to normalisations and charges (we take here an overall $-\pi$ factor with respect to the expression given in \cite{Kiritsis:2007zza}, even though we agree on the Fourier expansion). This expression can be motivated by considering the expected behaviour in $\ln |z|$ near the source, and further asking for invariance under the torus symmetries, which brings the $\theta_1$-function and the additional phase term.

We now proceed as in section \ref{sec:Green} to get the generalized Green's function \eqref{a3}, starting from \eqref{Fourier2d}: we obtain
\beq
G =  \int_0^\infty \d t \Big( 1 - \sum_{\vec{n}=(p,q) \in \mathbb{Z}^{2}} e^{2\pi \i \vec{n} \cdot \vec{\sigma} - 4\pi^2 |\tau p - q|^2 t} \Big) \ .
\eeq
Let us first consider $\tau_1=0$. In that case, we easily reconstruct a formula analogous to \eqref{a3}, namely
\beq
\tau_1=0:\quad G =  \int_0^\infty \d t \Big( 1 - \theta_3(\sigma^1|4\pi \i t \tau_2^2)\, \theta_3(\sigma^2|4\pi \i t) \Big) \ . \label{G2dtau10}
\eeq
For $\tau_1 \neq 0$, it is more difficult to separate into two $\theta_3$-functions. One can still rotate the vector $\vec{n}=(p,q)$ by an angle $\varphi$ towards $(\tilde{p}, \tilde{q})$, with
\beq
\cos^2 \varphi = \tfrac{1}{2} \left(1 + \tfrac{1 - |\tau|^2}{\sqrt{(1 + |\tau|^2)^2 - 4 \tau_2^2}} \right) \ ,
\eeq
such that $|\tau p - q|^2 = \lambda_1 \tilde{p}^2 + \lambda_2 \tilde{q}^2$. Doing the same rotation within $\vec{n} \cdot \vec{\sigma}$ then gives separability into a product of two sums on $\tilde{p}$ and $\tilde{q}$. But those are not necessarily integers, so we do not reconstruct the $\theta_3$-functions. We leave to future investigations this more general case.

\begin{figure}[H]
\begin{center}
\includegraphics[width=0.6\textwidth]{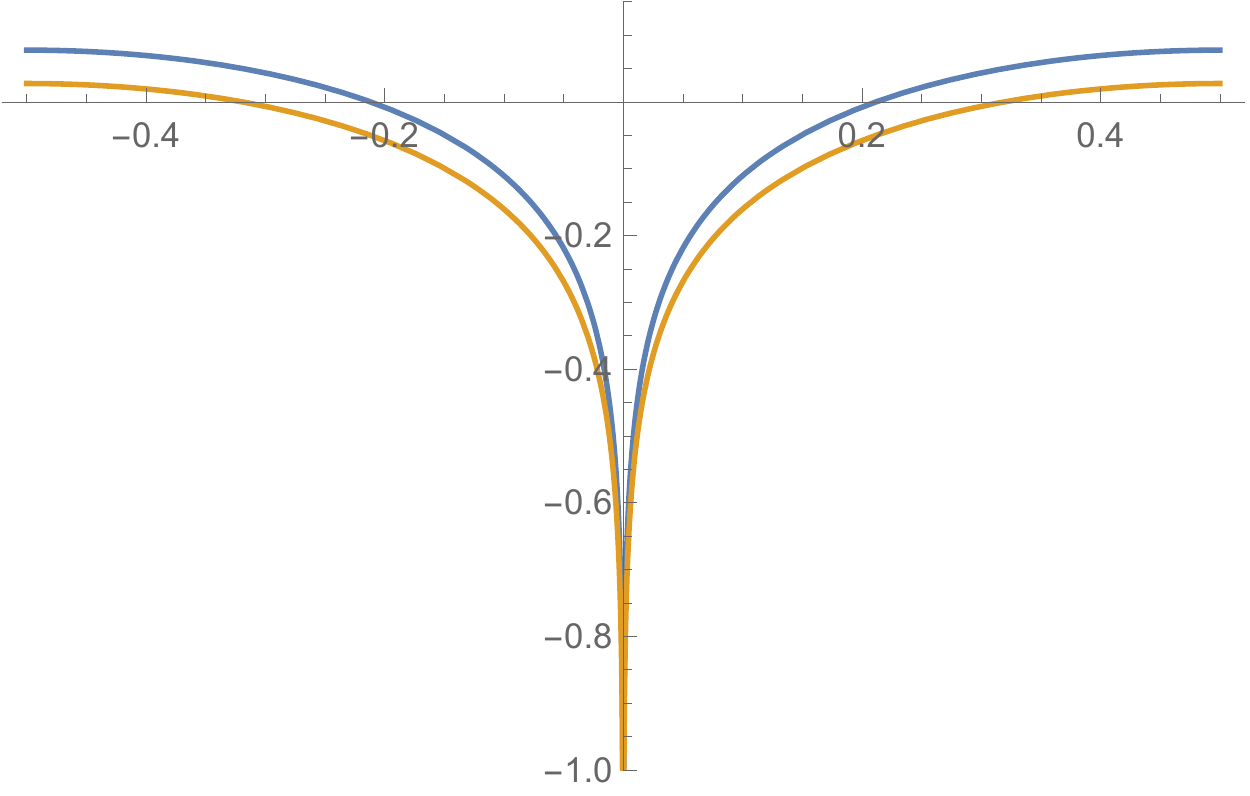}
\caption{Comparison of the graphs of the two Green's function expressions $\tilde{G}$ \eqref{G2d} and $G$ \eqref{G2dtau10} along $\vec{\sigma}=(\sigma^1,0)$, for $\tau=\i$. The lower orange curve is $G$ and the upper blue curve is $\tilde{G} - \tilde{G}(\sigma^1=\tfrac{1}{2}) + G(\sigma^1=\tfrac{1}{2}) + 0.05$. The constant between the functions is adjusted by the value at $\sigma^1=\tfrac{1}{2}$ and we add on the graph a shift of $0.05$ to be able to  see both curves; without the latter, they are simply indistinguishable. The same perfect match is observed along $\vec{\sigma}=(\sigma,\sigma)$ for which we do not display the graph here.}\label{fig:G2dcompare}
\end{center}
\end{figure}

The question is now whether, at least in the case $\tau_1=0$, the two expressions $\tilde{G}$ \eqref{G2d} and $G$ \eqref{G2dtau10} match. We started from the same Fourier expansion, but then proceeded with seemingly different regularizations. It would be interesting to study and compare both procedures in more depth. The alternative expression \eqref{G2dtau10} may then have applications beyond the present paper, within e.g.~string amplitudes. A first promising check is the comparison of the two graphs, which perfectly match, as displayed in Figure \ref{fig:G2dcompare}.

\section{Alternative numerical method for the $d=1$ spectrum}\label{ap:num}

In section \ref{sec:num}, we determine the first eigenvalues $\mu_N$ in equation \eqref{inttt8}. For $d=1$, we present here an alternative numerical method that goes back to the work of Hartree \cite{Hartree}, and was used for the determination of a Kaluza--Klein spectrum for the first time in \cite{Richard:2014qsa}. The idea goes as follows: the eigenvalue problem \eqref{inttt8} only has solutions for discrete values $\mu_N$ of $\mu$. Indeed, for an arbitrary $\mu$ we may obtain a unique solution $\psi$ (up to a normalisation by an overall constant) by imposing appropriate boundary conditions at $\sigma=0$. For the same $\mu$ we may obtain another  unique solution $\tilde{\psi}$ (up to normalisation) by imposing boundary conditions at $\sigma=\frac12$. The two solutions thus obtained will be independent unless their Wronskian vanishes
\eq{
W[\psi,\tilde{\psi}]= \psi\tilde{\psi}'-\tilde{\psi}\psi'=0~,
}
at some $\sigma_0\in[0,\frac12]$, in which case it vanishes identically for all $\sigma$. Let us now evaluate $W[\psi,\tilde{\psi}]$ at some fixed $\sigma_0$ and view it as a function of $\mu$. The Kaluza--Klein spectrum  will then be given by the  values $\mu=\mu_N$ for which the Wronskian vanishes.

\begin{figure}[H]
\begin{center}
\begin{subfigure}[H]{0.4\textwidth}
\includegraphics[width=\textwidth]{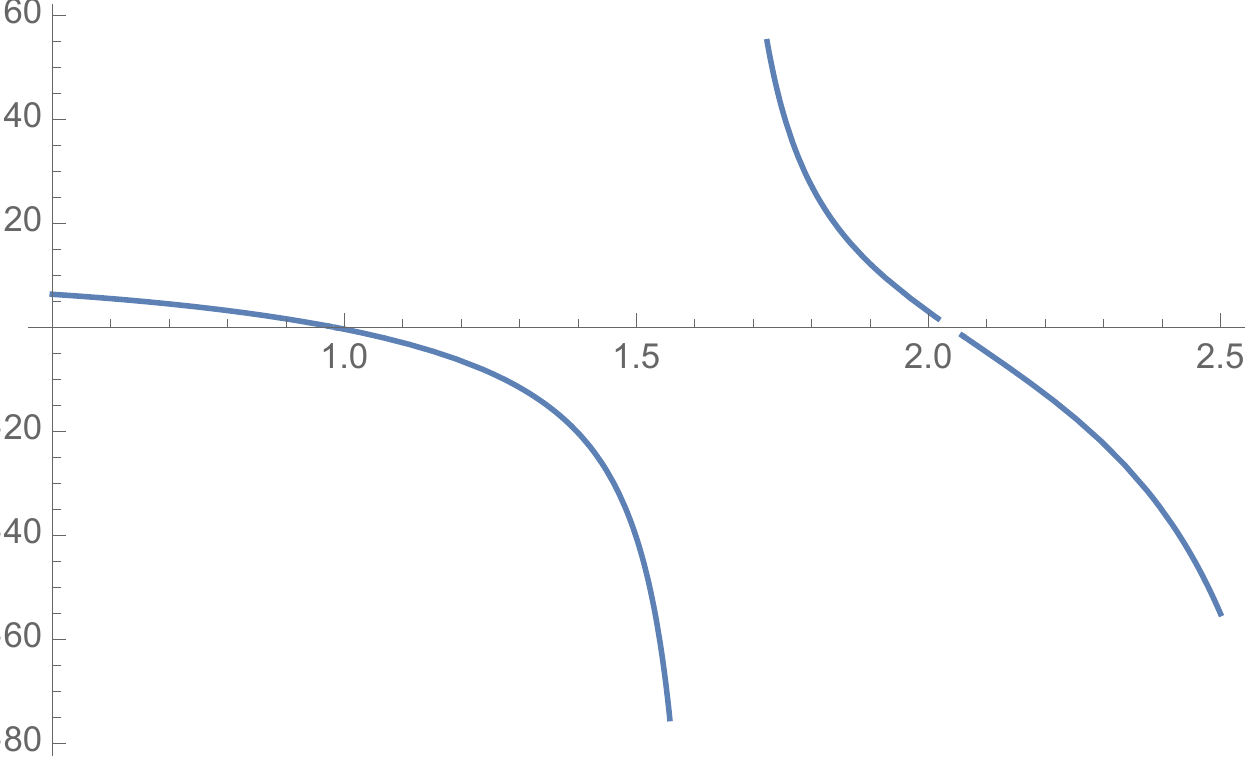}\caption{Odd functions}\label{fig:Wronskodd}
\end{subfigure}
\qquad \qquad
\begin{subfigure}[H]{0.4\textwidth}
\includegraphics[width=\textwidth]{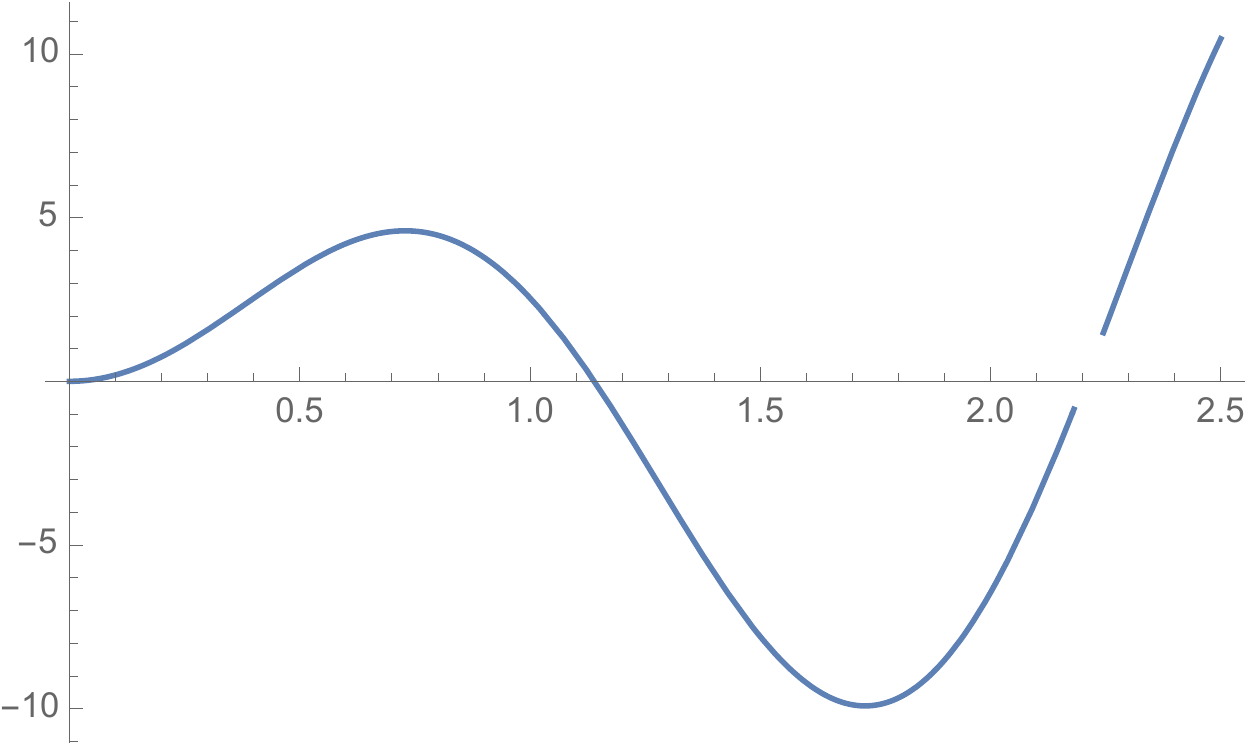}\caption{Even functions}\label{fig:Wronskeven}
\end{subfigure}
\caption{The Wronskian as a function of $\mu$ is depicted for odd or even functions. Its zeros $\mu=\mu_N$ give the spectrum of Table \ref{tab:specd=1}.}\label{fig:f3}
\end{center}
\end{figure}

The case at hand was first studied in section \ref{sec:normH}: the warp factor $H$ was given in \eqref{H1d}. The corresponding function $H'$ \eqref{hfnc} entering equation \eqref{inttt8} is given by
\eq{\label{419h}
H'=1-(2\pi L)^{-1}\left(8G(\sigma)-8G(\sigma-\tfrac12)
\right)
~,}
and turns out to be equal to the triangle function, as depicted in Figure \ref{fig:H1dall} and shown around \eqref{H1dFourier}. We then use the triangle function, and solve \eqref{inttt8} as explained above, imposing appropriate boundary conditions for the functions $\psi$. More precisely, at $\sigma=0,\tfrac12$ we have imposed $\psi(\sigma)=0$ or $\psi'(\sigma)=0$ for an odd or even function, respectively. We plot the Wronskian at $\sigma=0$ as a function of $\mu$, for odd functions on Figure \ref{fig:Wronskodd} and even ones on Figure \ref{fig:Wronskeven}. We find that the zeros of the Wronskian are in
agreement with the Kaluza--Klein spectrum given in Table \ref{tab:specd=1}.

\end{appendix}

\end{document}